\documentclass[twocolumn,english]{revtex4-2}
\usepackage[T1]{fontenc}
\usepackage[latin9]{inputenc}
\usepackage[a4paper]{geometry}
\usepackage{hyperref}
\usepackage{amssymb}
\usepackage{soul}
\usepackage{amsmath,amssymb,amsfonts}
\usepackage{comment}
\usepackage{environ}
\usepackage{braket}
\usepackage[normalem]{ulem}
\usepackage{tikz-feynman}
\usepackage{float}
\usepackage[dvipsnames]{xcolor}

\usetikzlibrary{shapes.misc}
\usetikzlibrary{decorations.markings}
\tikzset{
  midarrow/.style={
    postaction={decorate},
    decoration={markings, mark=at position .5 with {\arrow{>}}}
  }
}

\usepackage{wasysym}

\usepackage{bm}      
\usepackage{amsthm}  

\def\L{\Lambda}
\newcommand{\varep}{\varepsilon}
\newcommand{\nno}{\nonumber}
\newcommand{\tr}{\mbox{tr}}
\newcommand{\Tr}{\mbox{Tr}}

\NewEnviron{eqn}{%
  \begin{align}
    \begin{split}
      \BODY
    \end{split}
  \end{align}
}

\NewEnviron{eqn*}{%
  \begin{align*}
    \begin{split}
      \BODY
    \end{split}
  \end{align*}
}

\newcommand{\wb}{\bar w}

\newcommand{\intnsinf}{\int_{-\infty}^{0}}

\tikzset{
  cut mark/.style args={#1}{
    decoration={
      markings,
      mark=at position #1 with {
        \draw[red,line width=1.1pt]
          (0,-3pt) -- (0,3pt);
      }
    },
    postaction={decorate}
  }
}

\def\be{\begin{equation}}
\def\ee{\end{equation}}
\def\p{\partial}

\def\a{\alpha}
\def\b{\beta}
\def\k{\kappa}

\def\d{\delta}

\def\e{\epsilon}

\begin{document}
\widetext
\title{All Order Self-Dual Correlators From Symmetries}
\author{Chandramouli Chowdhury$^{a}$}
\email{chandra@itp.ac.cn}
\author{Silvia Nagy$^{b}$}
\email{silvia.nagy@durham.ac.uk}
\affiliation{$^{a}$New Cornerstone Laboratory, Institute of Theoretical Physics, Chinese Academy of Sciences, Beijing 100190, China,}
\affiliation{$^{b}$Department of Mathematical Sciences, Durham University, Durham, DH1 3LE, UK}

\begin{abstract}
We consider the scalar actions for self-dual Yang-Mills and self-dual gravity. We give a novel resummed version of their infinite tower of symmetries. We show that the Ward identities for symmetries corresponding to the individual levels of the infinite tower give the soft limits for wavefunction coefficients, while their resummed version completely fixes the full wavefunction. Our analysis is valid for general boundary conditions. Finally, we give an explicit formula for the n-point wavefunction coefficient as a single determinant reorganising the perturbiner solution, which also satisfies the soft limits.

\end{abstract}
 \maketitle
\makeatletter
\begingroup
\def\l@subsection#1#2{}
\def\l@subsubsection#1#2{}
\def\l@paragraph#1#2{}
\def\l@subparagraph#1#2{}
\tableofcontents
\endgroup
\makeatother
\twocolumngrid

\section{Introduction}\label{sec:intro}
The self-dual subsectors of Yang-Mills and gravity has been studied extensively due to the dramatic simplifications that occur in this context. This provides a toy model which allows for real progress on various proposals, such as the double copy \cite{Monteiro:2011pc,Chen:2019ywi,Chen:2021chy,Monteiro:2022nqt,Monteiro:2013rya,Elor:2020nqe,Armstrong-Williams:2022apo,Farnsworth:2021wvs,Skvortsov:2022unu,Campiglia:2021srh,Ben-Shahar:2022ixa,CarrilloGonzalez:2024sto,Berman:2018hwd,Nagy:2022xxs,Monteiro:2022lwm,Krasnov:2021cva,Monteiro:2022xwq,Lipstein:2023pih,Borsten:2023paw,Brown:2023zxm,Doran:2023cmj,Chowdhury:2024dcy,LopesCardoso:2024ttc,Correa:2024mub,Diaz-Jaramillo:2025gxw,LopesCardoso:2025ldi}, and, more recently, holography \cite{Skvortsov:2026ofl, Skvortsov:2026dru, Skvortsov:2026gtq}.  

Tree level amplitudes vanish in this setting (with the exception of recently studied exceptions in the half-collinear regime \cite{Guevara:2026qzd, Guevara:2026qwa}) and it is one loop exact. While this exhausts the list of asymptotic physical observables,  there are non-vanishing tree-level observables such as correlation functions defined at finite times. These are particularly useful in curved backgrounds. For this purpose, it is often convenient to first study the wavefunction \cite{Hartle:1983ai, Hatfield:1992rz, Maldacena:2002vr}, and through that, obtain the correlators.

The self dual sector has also been studied as an integrable model and is known to possess an infinite number of symmetries \cite{Plebanski:1975wn,Prasad:1979zc,Dolan:1983bp,Popov:1996uu,Popov:1998pc,Park:1989vq,Parkes:1992rz,Husain:1993dp,Husain:1994vi,Bardeen:1995gk,Chalmers:1996rq,Cangemi:1996pf,Wolf:2004hp,Popov:2006qu,Papachristou_2021,Adamo:2021lrv} defined recursively (indeed the non-vanishing 1-loop amplitude can be interpreted as an anomaly thereof \cite{Bardeen:1995gk,  Monteiro:2022nqt, Doran:2023cmj, Costello:2021bah}). In \cite{Campiglia:2021srh, Nagy:2022xxs}, the first few steps were taken towards solving the recursion relations and interpreting these symmetries as asymptotic symmetries related to soft theorems.  

In this article we revisit this infinite tower of symmetries, and show that they can be resumed as a single-parameter symmetry. We find that the Ward identity corresponding to this symmetry \emph{uniquely} determines the tree-level Bunch-Davies wavefunction at arbitrary multiplicity. The resulting coefficients agree with those obtained from the perturbiner solution. Our analysis is valid for wavefunctions with Dirichlet, Neumann, or Robin boundary conditions.

This distinguishes them from ordinary gauge Ward identities, which generally leave dynamical information undetermined, but give consistency conditions such as the soft limits \cite{Creminelli:2004yq, 
Hui:2018cag, Pimentel:2013gza, Bittermann:2022nfh, McFadden:2014nta,  Banerjee:2021llh, Hamada:2018vrw, Chowdhury:2024wwe, Hinterbichler:2013dpa, Chowdhury:2024snc, Albayrak:2024ddg, Creminelli:2012ed}. We also recover such soft limits in our theory through the Ward identities corresponding to individual families in the infinite tower, which organises the symmetries corresponding to all subleading orders in the soft expansion. At first order we recover an Adler zero. The constraints apply as one of the components of the momentum vanishes, without requiring the remaining independent momentum components to vanish, and therefore probe more general kinematics than the usual full soft limit. As a special case, we also recover the diagrammatic soft limits previously computed in \cite{Chowdhury:2024dcy}.   

Finally, we express the wavefunction coefficients obtained from the perturbiner as a determinant of an (almost) upper-triangular matrix, which we explicitly provide for SDYM. This does not exhibit a total energy pole, and gives the expected vanishing amplitudes in the flat space limit. This organizes the recursive structure of the perturbiner, but does not establish a substantial computational simplification as its size grows exponentially with multiplicity (also see recent works containing all multiplicity recursive formulas for wavefunctions in different theories \cite{Gomez:2026yno, Heuveline:2026knw, He:2026sja, Nabavi:2026zyk}). 


The article is organised as follows. In \autoref{sec:review} we review the scalar descriptions of self-dual Yang-Mills (SDYM) and self-dual gravity (SDG), and explain the resummation of the infinite towers of symmetries. In \autoref{sec:classicalanomaly} we compute the charges associated with these symmetries. In \autoref{sec:wavef} we show that the Ward identity for these charges allows us to fully reconstruct the wavefunction, and the answer agrees with the perturbiner calculation. \autoref{sec:familysymmsoft} derives the soft limits for the wavefunction coefficients from the family-by-family Ward identities. In \autoref{sec:nptPsi} we give the explicit $n$-point wavefunction coefficient as a determinant formula. We conclude in \autoref{sec:conc}. We give more details of the symmetry calculations and wavefunction calculation in \autoref{Symmetry_calc_app} and \autoref{app:Witten}, respectively.

\section{Resummed symmetries of the self-dual sector}\label{sec:review}
We work in light-cone coordinates in Lorentzian signature:
$u{=}\tfrac{1}{\sqrt{2}}\left(t-z\right)$, $v{=}\tfrac{1}{\sqrt{2}}\left(t+z\right)$, $w{=}\tfrac{1}{\sqrt{2}}\left(x+iy\right)$, and $\bar{w}{=}\tfrac{1}{\sqrt{2}}\left(x-iy\right)$. We will find it useful to divide our spacetime into two subplanes, spanned by coordinates
\be \label{plane_splitting}
x^i = (u, \wb), \qquad y^\a = (v, w)
\ee 
Then the line element becomes
\be\label{LC_eta_def} 
ds^2=2\eta_{i\alpha}dx^idy^\alpha =-2dudv+2dwd\bar{w}
\ee 
\subsection{Self-dual YM}
\label{subsec:SDYMsymm}
For YM, the self-duality equation
\be\label{SD_YM_def}
F_{\mu\nu}=\tfrac{\sqrt{g}}{2}\epsilon_{\mu\nu\rho\sigma}F^{\rho\sigma} \qquad \Leftrightarrow \qquad F_{\mu\nu}^-=0\, , 
\ee 
reduces, in light-cone gauge $A_u=0$ to the scalar equation
\be\label{Phi_eom} 
\square\Phi+ig\varepsilon^{ij} \left[\partial_i\Phi,\partial_j\Phi \right]=0 \, ,
\ee 
with $\varepsilon^{\bar w u} =1$ and $\square=2\eta^{i\alpha}\partial_\alpha\partial_i=2\p^i\p_i$ and 
\be
\label{A_phi}
A_i=0,\qquad A_\alpha=\left(A_v,A_w\right)=\left(\partial_{\bar{w}}\Phi,\partial_u\Phi\right)\, .
\ee 
Due to the integrability of the SD sector \cite{Plebanski:1975wn,Prasad:1979zc,Dolan:1983bp,Popov:1996uu,Popov:1998pc,Park:1989vq,Parkes:1992rz,Husain:1993dp,Husain:1994vi,Bardeen:1995gk,Chalmers:1996rq,Cangemi:1996pf,Wolf:2004hp,Popov:2006qu,Papachristou_2021}, the scalar e.o.m. \eqref{Phi_eom} admits an infinite set of symmetries, obtained via the on-shell recursions 

\be
\partial_i\delta_{\Lambda_{n+1}}\Phi=\varepsilon_{ij}\p^j  \delta_{\Lambda_n}\Phi-ig\left[\partial_i\Phi,\delta_{\Lambda_n}\Phi \right]
\ee
For our purposes, it will be sufficient to solve the weaker recursion, obtained by contracting the above with $2\p^i$

\be 
\square\delta_{\Lambda_{n+1}}\Phi=-ig\left[\square\Phi,\delta_{\Lambda_n}\Phi \right]
-2ig\left[\partial_i\Phi,\partial^i\delta_{\Lambda_n}\Phi \right]
\ee 
Start with the following seed symmetry of the e.o.m. \eqref{Phi_eom}, which we will call the $0$th family:
\be \label{0th}
\delta_{\Lambda_0}\Phi=\bar{\Lambda}_0(y)
\ee 
We can solve the recursions perturbatively to obtain an infinite family of symmetries
\begin{eqn}
\delta_{\Lambda_n}\Phi=&\frac{1}{n}\left(x^i\varepsilon_{ij}\partial^j\right)\Lambda_n+ig\left[\Lambda_n,\Phi\right]\\
&+2ig\frac{1}{\square}\left[\partial^i\Phi,\partial_i\Lambda_n\right]+\mathcal{O}\left(g^2 \right)
\end{eqn}
where
\be \label{LAmbda_n_def}
\Lambda_n(x, y) =\frac{1}{(n-1)!}\left(x^i\varepsilon_{ij}\partial^j\right)^{n-1}\bar\Lambda_n(y)\, ,
\ee 
and 
\be
\frac{1}{\square}f(x)=\int d^4y G(x,y)f(y),\quad  \square G (x,y)=\delta^{(4)}(x-y)
\ee 
with $\bar\Lambda_n$ an arbitrary function of $y$. We remark that the first family ($n=1$) is a residual gauge symmetry of the YM theory, with $\Lambda_1(x, y)=\Lambda_1(y)$, whereas the higher families are only symmetries of the SD sector. 

We can sum over all the families to write a single transformation
\be \label{resummed_symm}
\delta_{\Lambda}\Phi=\tilde{\Lambda}+ig\left[\Lambda,\Phi\right]
+2ig\frac{1}{\square}\left[\partial^i\Phi,\partial_i\Lambda\right]+\mathcal{O}\left(g^2 \right)
\, ,
\ee
where $\L(x,y)$ and and $\tilde\L(x,y)$ are not independent parameters, but related via
\be \label{LtildeL}
\varepsilon_{ij} \p^j \Lambda = \p_i \tilde \Lambda
\ee 
and are both general harmonic functions
\be \label{harmonic}
\square \tilde\Lambda =0=\square \L
\ee 
as shown in \autoref{App:Symmetry parameter}. The extension of \eqref{resummed_symm} to all orders in $g$ is given in \autoref{app:allorders}. The e.o.m. in \eqref{Phi_eom}  arises from the Lagrangian
\be\label{scalar_Lag} 
S_{YM}=\int d^4 x\ \text{Tr}\ \bar{\Phi}\left(\square\Phi+ig\varepsilon^{ij}\left[\partial_i\Phi,\partial_j\Phi \right]\right) \,
\ee
which is a truncation of the full YM Lagrangian \cite{Chalmers:1996rq,Chowdhury:2024dcy}. Requiring a well defined variational principle with boundary values of the field fixed, we find that we need the boundary term 
\be \label{SYMDirBndy}
S_{YM}^{\p D}=-\int d^4 x\ \text{Tr}\partial_\mu \left(\bar\Phi\partial^\mu \Phi \right) 
\ee 
For convenience we denote $S_D \equiv S_{YM}+S_{YM}^{\p D}$. Our results easily generalize to other boundary conditions as shown in \autoref{sec:wavef}.

We find that the Lagrangian and the e.o.m. (for both $\Phi$ and $\bar\Phi$) preserve the resummed symmetry if we supplement \eqref{resummed_symm} with 
\be \label{d_bar_phi_o_1}
\delta_\Lambda\bar{\Phi}=ig[\Lambda,\bar\Phi]
+2ig\frac{1}{\square}\left[\partial^i\bar\Phi_{\mathfrak{h}},\partial_i\Lambda\right] +\mathcal{O}\left(g^2 \right)
\ee 
where we introduced the notation $f_{\mathfrak{h}}=f-\frac{1}{\square}\square f$, denoting the harmonic part of $f$, i.e. $\square f_{\mathfrak{h}}=0$. For expressions at all orders in $g$, see \autoref{app:allorders}. The boundary term in \eqref{SYMDirBndy} gives a contribution to the charge as will be shown in \autoref{sec:classicalanomaly}.

\subsection{Self-Dual GR}
There is a similar construction for the action in the self-dual sector of Einstein GR \cite{}.
\be\label{scalar_Lag_GR} 
S_{GR}=\int d^4 x\ \bar{\phi}\left(\square\phi-\frac\kappa2 \varepsilon^{ij}\{\partial_i\phi,\partial_j\phi\}\right) \,
\ee
where $\k$ is the coupling constant and the Poisson bracket is defined as
\begin{eqn}
\{f, g\} = \varepsilon^{ij} \p_{i} f \p_j g
\end{eqn}
Remarkably we can obtain the symmetries for the gravity case via double copy from YM (first discovered at the level of the e.o.m. in \cite{Monteiro:2011pc}).
\begin{eqn}\label{DCrules}
\Phi\to \phi,\quad 
-ig[,]\to \k\frac{1}{2}\{,\},\quad
\Lambda\to\lambda
\end{eqn}
Using this prescription the symmetries are given as,
\begin{eqn}
\delta_{\lambda}\phi&=\tilde{\lambda}-\frac\kappa2 \{\lambda,\phi\}
-\k\frac{1}{\square}\{\partial^i\phi,\partial_i\lambda\}+\mathcal{O}\left(\k^2 \right), \\
\delta_\lambda\bar{\phi}&=-\frac\k2\{\lambda,\bar\phi\}
-\kappa\frac{1}{\square}\{\partial^i\bar\phi_{\mathfrak{h}},\partial_i\lambda\} +\mathcal{O}\left(\k^2 \right)
\end{eqn}
For the interested reader, the appropriate boundary term for a well defined variation with Dirichlet-Dirichlet boundary condition is given in \eqref{SDGbndy}.

\section{Charges}\label{sec:classicalanomaly}
The classical conserved charge for the symmetry in the \autoref{subsec:SDYMsymm} is given by
\be
Q_\L= \Tr\int_{\Sigma_0}d^3x \left(\left(\dot{\bar{\Phi}}-i\sqrt{2}g[\p_{\bar{w}}\Phi,\bar{\Phi}]\right)\d_\L\Phi- \bar{\Phi}\p_t\d_\L\Phi \right)~.
\ee
We now compute the Noether charges corresponding to the symmetries using the general formula
\begin{eqn}\label{NoetherGen}
Q_\lambda = \int_{\Sigma_0} \sum_{i} \pi_{\mathcal F_i} \delta_\lambda \mathcal F_i - \delta_\lambda S(\{\mathcal F\}) 
\end{eqn}
where $\mathcal F_i$ denotes all the fields appearing in the system and $\pi_{\mathcal F_i}$ is the conjugate momentum corresponding to those fields. The charge is written on the $t = 0$ hypersurface denoted by $\Sigma_{0}$ with the standard Minkowski metric given in \eqref{timemetric}.  The analysis is similar for the SDYM and SDG case. This is because the relevant boundary terms to be added for using appropriate boundary conditions are similar for  SDYM and SDG. Hence we carry out the analysis explicitly for the SDYM case with the SDG case following in an analogous manner. 

A well defined conjugate momentum in \eqref{NoetherGen} is obtained from the full action \eqref{scalar_Lag} which includes the boundary term given in \eqref{SYMDirBndy} \footnote{Note that the boundary term for gravity is cubic in the field due to the extra derivatives appearing in the interaction term of the Lagrangian.},
\begin{eqn} \label{conjmomYM}
\pi_\Phi&=\dot{\bar{\Phi}}-i\sqrt{2}g[\p_{\bar{w}}\Phi,\bar{\Phi}]\\
\pi_{\bar{\Phi}}&=\dot{\Phi}
\end{eqn}
The on-shell variation of the action appearing in \eqref{NoetherGen} only receives a contribution from the boundary term \eqref{SYMDirBndy} which is given as 
\begin{eqn}\label{Sdvar}
\delta_\Lambda S_{YM}^{\p D}  = \Tr \int_{\Sigma_0} d^3x \Big( \delta_\Lambda \bar\Phi \p_t \Phi +  \bar\Phi \p_t \delta_\Lambda \Phi \Big)
\end{eqn}
Together with \eqref{conjmomYM}  we see that the charge \eqref{NoetherGen} can be expressed in a simplified form
\begin{eqn}\label{eq:chargegen}
Q_\Lambda &= \int_{\Sigma_0}d^3x \pi_{\Phi} \delta_\Lambda \Phi + \pi_{\bar\Phi} \delta_\Lambda \bar\Phi - \delta_\Lambda S_{YM}^{\p D}  \\
&= \Tr\int_{\Sigma_0}d^3x \left(\pi_\Phi\d_\L\Phi- \bar{\Phi}\p_t\d_\L\Phi \right)~.
\end{eqn}
The charge is independent of the variation of $\bar\Phi$. We now promote the charge to a quantum operator. With Dirichlet boundary condition the appropriate replacement for the conjugate momentum is  $\pi_{\Phi}\to -i\frac{\delta}{\delta\Phi}$ and $\pi_{\bar\Phi}\to -i\frac{\delta}{\delta\bar\Phi}$, with the second to be performed after evaluating $\p_t\d_\L\Phi$ explicitly. We also choose a particular operator ordering which gives the following expression for the charge \eqref{eq:chargegen}
\begin{eqn}\label{charge_Dirichlet}
\widehat Q_\Lambda^D = \Tr\int_{\Sigma_0}d^3x  \left(-i \d_\L\Phi \frac{\delta}{\delta \Phi(\vec x)} - \bar{\Phi}\p_t\d_\L\Phi \right)
\end{eqn}
In section \ref{ssec:WFbndycond} we demonstrate the implication of this on the ward identity for the wavefunctions in the appropriate basis.

\section{Uniqueness of the Wavefunction}\label{sec:wavef}
We now explain how the symmetry transformations discussed above affect the ground state wavefunction of the theories. Conventionally, wavefunctions can be evaluated using Witten diagrams \cite{Chowdhury:2024dcy, Skvortsov:2026dru, Skvortsov:2026ofl, Skvortsov:2026gtq} as reviewed in appendix \ref{app:Witten}. However in this section we demonstrate how these can be recursively evaluated by solving the Ward identities corresponding to the self-dual symmetries. Unlike the standard ward identities arising from the gauge (and conformal symmetry) \cite{Berezhiani:2013ewa, Pimentel:2013gza, Kundu:2015xta, Ghosh:2014kba, Chowdhury:2021nxw, Bzowski:2013sza,  Hinterbichler:2013dpa, Arkani-Hamed:2018kmz, Baumann:2020dch, Chen:2025ljl} we show how the self dual symmetries {\it completely} fix the tree level wavefunction coefficients at all points recursively. 
 
For the SDYM case the flat space description is equally valid for the wavefunction in any conformally flat spacetime, such as dS (or analytically continued correlators in AdS) \cite{Chowdhury:2024dcy, CarrilloGonzalez:2024sto, Skvortsov:2026gtq, Skvortsov:2026ofl, Heuveline:2026knw}. In flat space the formulas are structurally the same for both SDYM and SDG and hence we phrase the discussion explicitly for SDYM.

The line element convenient for this description is given as
\be\label{timemetric}
ds^2=dz^2-dt^2+2dw\,d\wb = dz^2 - dt^2 + dx^2 + dy^2~.
\ee
The ground state Dirichlet wavefunction is obtained by a path integral over
$t \leq 0$, with fixed field values at $t = 0$ and regularity as $t\to-\infty(1 - i \e)$. This regularity condition is often known as the Bunch-Davies boundary condition which is explicitly given as
\be \label{BD}
\Phi(-\infty(1 - i \e)) = 0\, ,\quad 
\bar\Phi(-\infty(1 - i \e)) = 0 \, ,
\ee
With these conditions the path integral definition of the wavefunction is
\begin{eqn}\label{eq:WFdefn}
\Psi[\varphi, \bar\varphi] = \int_{\rm BD}^{\substack{
  \Phi(0)=\varphi\\
  \bar\Phi(0)=\bar\varphi
}} D\Phi D\bar\Phi e^{i S_{YM}[\Phi, \bar\Phi] + i S_{YM}^{\p D}[\Phi, \bar\Phi]}
\end{eqn}
 The upper limit of the integral fixes the values of the field at $t = 0$. We comment on the wavefunctions obtained by fixing Neumann and Robin boundary condition in a later section \ref{ssec:WFbndycond}. There is a special choice for the boundary conditions that results in a wavefunction proportional to a delta function given in \ref{ssec:deltaWF}.

For a momentum $\vec k$ along the $t=0$ slice, we use the notation $k=|\vec k|$. These are the three-dimensional momenta corresponding to the $(x, y, z)-$directions. For later use we also define a $u-$component of the momentum $k_u = \frac{1}{2}(k - k_z)$.

The wavefunctional at tree-level admits an expansion in field basis
\begin{eqn}\label{eq:WFexp}
\Psi[\varphi, \bar\varphi] &= e^{\sum_{n = 2}^\infty g^{n-2} \mathcal W_n[\varphi, \bar\varphi]}
\end{eqn}
where $\mathcal W_n$ are expressed in terms of the wavefunction coefficients,
\begin{eqn}
\mathcal W_n = \Tr \int_{12\cdots n}   \psi_n(\vec k_1, \cdots, \vec k_n)  \varphi(\vec k_1) \cdots \varphi(\vec k_{n-1}) \bar \varphi(\vec k_n)
\label{eq:wavefunction_expansion}
\end{eqn}
where the notation $\int_{12\cdots n} \equiv \int d^3 k_1 d^3 k_2 \cdots d^3 k_n \delta(\vec k_1 + \vec k_2 + \cdots + \vec k_n)$ and $\psi_n(\vec k_1, \cdots, \vec k_n)$ are known as the wavefunction coefficients. The Tr denotes the trace taken over the color index. This representation is analogous to the single-minus gluon wavefunction.

\subsection{Ward Identity from Classical Anomaly}
We now demonstrate how the symmetry transformations affect the wavefunctions and give the Ward identities corresponding to these symmetries. Since we want to keep the BD vacuum invariant under the symmetry transformations, the parameter $\Lambda$ should preserve these boundary conditions. Along with the harmonic condition \eqref{harmonic} this automatically fixes
\begin{eqn}\label{gauge_falloff}
\tilde{\L}(t,\vec{k})=\Lambda_0(\vec k)e^{+itk}\quad\Rightarrow\quad \dot{\tilde\L}(t, \vec k)=ik \tilde \Lambda(t, \vec k) ~.
\end{eqn} 
To obtain the ward identity we first evaluate the transformation of the wavefunction under a variation of the fields
\begin{eqn}\label{psitrans}
&\Psi(\varphi + \delta_\Lambda \varphi, \bar\varphi + \delta_\Lambda \bar\varphi)\\
&= \Psi(\varphi, \bar\varphi) + i \int d^3 x \Big( \delta_\L \varphi \pi_{\varphi} + \delta_\L \bar\varphi \pi_{\bar\varphi} \Big) \Psi(\varphi, \bar\varphi)
\end{eqn}
where we have suppressed terms of $O(\Lambda^2)$ and recognize $\pi_{\varphi} = -i \frac{\delta}{\delta \varphi}, \pi_{\bar\varphi} = -i \frac{\delta}{\delta \bar\varphi}$. Using \eqref{eq:chargegen} the equation above can also be written as
\begin{eqn}\label{psitrans2Q}
&\Psi(\varphi + \delta_\Lambda \varphi, \bar\varphi + \delta_\Lambda \bar\varphi) = \Psi(\varphi, \bar\varphi) + i \Big( \widehat Q_\Lambda + \delta_\Lambda S_{YM}^{\p D} \Big) \Psi(\varphi, \bar\varphi)
\end{eqn}
We now provide an alternate expression for the variation of the wavefunction that recovers the conventional form of the ward identity. From the definition \eqref{psitrans} we have,
\begin{eqn}
\Psi(\varphi + \delta_\Lambda\varphi,
     \bar\varphi + \delta_\Lambda\bar\varphi)
=
\int_{\mathrm{BD}}^{\substack{
  \Phi(0)=\varphi+\delta_\Lambda\varphi\\
  \bar\Phi(0)=\bar\varphi+\delta_\Lambda\bar\varphi
}}
D\Phi\,D\bar\Phi\,
e^{iS_D[\Phi,\bar\Phi]} .
\end{eqn}
We perform a change of variables in the path integral 
\begin{eqn}
\Phi' = \Phi - \delta_\Lambda \Phi, \qquad \bar \Phi' = \bar\Phi - \delta_\Lambda \bar\Phi
\end{eqn}
which leads to
\begin{align}
&\Psi(\varphi+\delta_\Lambda\varphi,
      \bar\varphi+\delta_\Lambda\bar\varphi)\nno \\
&=
\int_{\mathrm{BD}}^{\substack{
  \Phi'(0)=\varphi\\
  \bar\Phi'(0)=\bar\varphi
}}
D\Phi'\,D\bar\Phi'\,
e^{iS_D[\Phi'+\delta_\Lambda\Phi,\,
\bar\Phi'+\delta_\Lambda\bar\Phi]}
\\
&=
\int_{\mathrm{BD}}^{\substack{
  \Phi(0)=\varphi\\
  \bar\Phi(0)=\bar\varphi
}}
D\Phi\,D\bar\Phi\,
e^{iS_D[\Phi,\bar\Phi]}
\Bigl(1+i\delta_\Lambda S_D[\Phi,\bar\Phi] +O(\Lambda^2)\Bigr) \nno\\
&\equiv
\left[1+i \delta_\Lambda S_{YM}^{\p D} +O(\Lambda^2)\right]
\psi(\varphi,\bar\varphi)
\nno
\end{align}
In the second equality we have used the gauge invariance of the BD vacuum.  
Since the bulk action is invariant under the gauge transformation and the only change is in the boundary term we have also used $ \delta_\Lambda S_D =  \delta_\Lambda S_{YM}^{\p D}$ in the final equality. Note that we have ignored quantum anomalies that arise in the transformation of the measure since we are only interested in tree level wavefunctions \cite{, Monteiro:2022nqt}. Thus the transformation law reads
\begin{eqn}\label{WFvargen}
\Psi(\varphi+\delta_\Lambda\varphi,\bar\varphi+\delta_\Lambda\bar\varphi)
=
\Psi(\varphi, \bar\varphi) + i \delta_\Lambda S_{YM}^{\p D}
\Psi(\varphi,\bar\varphi).
\end{eqn}
 We refer to this as the \emph{classical anomaly equation} reflecting the non-trivial transformation of the boundary action. One could extend this to loop level by adapting the computation done in scattering amplitudes to wavefunctions \cite{Doran:2023cmj}. Such transformations of the wavefunctions also appear in theories with non-trivial topological properties such as Chern-Simons theories \cite{Arcioni:2002vv}.

Comparing \eqref{psitrans2Q} and \eqref{WFvargen} we immediately recover the conventional form of the Ward identity for the wavefunctional.
\begin{eqn}\label{genQPhi0}
\widehat{Q}_\Lambda \Psi =0
\end{eqn}
We use this identity in two ways. Expanding the resummed symmetry order by order in $g$ determines the full coefficient $\psi_n$ at order $g^{n-2}$, as shown below. Applying the Ward identities to individual symmetry families instead constrains the expansion in the $p_{\bar w}$ component of the momentum of a $\varphi$ leg (see section \ref{sec:familysymmsoft}). 

\subsection{2 Points}\label{sec:2ptPsiproof}
We first demonstrate the implication of the ward identity for the 2-pt  wavefunction coefficient. For this it suffices to consider the charge at $O(g^0)$ which is equivalent to the charge in the linearized theory. From \eqref{charge_Dirichlet} we have
\begin{eqn}\label{QLambda0}
\widehat{Q}^{(0)}_{\Lambda_0}=& -\int d^3 p \text{Tr} 
\left[\bar{\varphi}(-\vec p)\dot{\tilde\L}_0(\vec p)+i\tilde\L_0(\vec p)\frac{\delta}{\delta\varphi(\vec p)}\right]
\end{eqn}
Then, at $O(g^0)$ from \eqref{genQPhi0}, using the gauge parameter fall-off \eqref{gauge_falloff}, we get
\begin{eqn}\label{Q0eW2=0}
  \widehat{Q}_\Lambda^{(0)} e^{\mathcal{W}_2}=
  -\int_{\Sigma_0}d^3p \text{Tr} 
\bar{\varphi}(-\vec p)\tilde\Lambda_0(\vec p)\left[ip  +i\psi_2\right] e^{\mathcal{W}_2}=0
\end{eqn}
from which we read off $\psi_2(\vec p) = - p$,
recovering the standard 2-pt wavefunction arising from the on-shell action. This method can be extended to perturbatively solve for the higher point wavefunction coefficients \footnote{We have explicitly checked this to 5-pt}. Instead of demonstrating this explicitly we show below that the Ward identity corresponding to the the resummed symmetry uniquely determines the full wavefunction which agrees with the perturbiner solution.

\subsection{$n$-point Perturbiner from the Ward identity}\label{sec:pert}
We now prove that the $n$-pt tree level wavefunction can be obtained recursively by solving the Ward identity. This is shown in two steps. In the first step we demonstrate that the perturbiner solution solves the ward identity. In the second step we prove that it is the unique solution.

The perturbiner method requires us to solve for the e.o.m for the field and then plug that into the on-shell action. The field solves the following equation of motion,
\begin{eqn}\label{perteq}
\Phi_{cl} = \Phi_0 - ig \frac{1}{\square} \varepsilon^{ij} [\p_i \Phi_{cl}, \p_j \Phi_{cl}]
\end{eqn}
where $\square \Phi_0 = 0$ is the solution for the free field and the Green function prescription needs to be specified for the $\frac{1}{\square} = - i G_D$ and it must be kept consistent, which for our case is the Dirichlet Green function (where $G_D$ is given in eq \eqref{Dgreenfunction} for an explicit expression). To compute $\Psi  =e^{\mathcal W}$, we need to evaluate the on-shell action,
\begin{eqn}\label{Wonshell}
\mathcal W = i (S_D)_{on-shell} = \left. i \Tr \int d^3 x \bar\varphi \p_t \Phi_{cl} \right|_{t =0}
\end{eqn}
where we have used \eqref{SYMDirBndy}. 
The classical solution $\Phi_{cl}[\varphi(t, \vec x)]$ can be obtained perturbatively as shown in more detail in section \ref{sec:nptPsi}. 
To make contact with the Ward identity \eqref{genQPhi0} we also evaluate its variational derivative below
\begin{eqn}\label{Wvar1}
&\Tr \int d^3 x \delta_\Lambda \varphi \frac{\delta\mathcal  W}{\delta \varphi} \\
&= \left.i \Tr \int d^3 x \delta_\Lambda \varphi(\vec x) \int d^3 y \bar\varphi(\vec y) \p_t \frac{\delta \Phi_{cl}[\varphi(t, \vec y)]}{\delta \varphi(\vec x)} \right|_{t = 0}~.
\end{eqn}
Since $\delta_\Lambda\varphi(\vec x)$ is a variation of the boundary data on the fixed slice $t=0$, it is held fixed when $\partial_t$ acts on the bulk solution. We may therefore move $\partial_t$ outside the integral to obtain
\begin{eqn}\label{deltadWdphi}
\text{\eqref{Wvar1}}
&= \left. i \Tr \int d^3 y \bar\varphi(\vec y) \p_t \int d^3 x \delta_\Lambda \varphi(\vec x) \frac{\delta \Phi_{cl}[\varphi(t, \vec y)]}{\delta \varphi(\vec x)} \right|_{t = 0}
\end{eqn}
The RHS can be simplified by using 
\begin{eqn}
 \int d^3 x \delta_\Lambda \varphi(\vec x) \frac{\delta \Phi_{cl}[\varphi(t, \vec y)]}{\delta \varphi(\vec x)}   = \delta_\Lambda\Phi_{cl}[\varphi(t, \vec y)]
\end{eqn}
which is the variation of the classical solution upon changing its boundary value. Hence \eqref{Wvar1} simplifies to 
\begin{eqn}
\Tr \int d^3 x \delta_\Lambda \varphi \frac{\delta \mathcal W}{\delta \varphi} &= \left. i \Tr \int d^3 y \bar\varphi(\vec y) \p_t \delta_{\Lambda} \Phi_{cl}[\varphi(t, \vec y)] \right|_{t = 0}
\end{eqn}
This proves that on-shell value of $\mathcal W$ solves the equation $\widehat Q_{\Lambda}^D \Psi= 0$ with $\widehat Q_{\Lambda}^D$ defined in \eqref{charge_Dirichlet}. 

We now prove the uniqueness of the solution. We prove this by contradiction. Consider another solution for the Ward identity and call it $\Psi_* = e^{\mathcal W_*}$. The two solutions $\mathcal W$ and $\mathcal W_*$ have the same quadratic term (as shown in \eqref{sec:2ptPsiproof}) and assume first differ at $O(g^{n_* -2})$ for $n_* \geq 3$. Thus 
\begin{eqn}
\mathcal W - \mathcal W_* 
&= g^{n_* -2} (\mathcal W_{n_*} - \mathcal W_{n_*}^*) + O(g^{n_* -1})
\end{eqn}
From this we obtain the difference of $\Psi$ as
\begin{eqn}
\Psi - \Psi_* = g^{n_* - 2} e^{\mathcal W_2} (\mathcal W_{n_*} - \mathcal W_{n_*}^*) + O(g^{n_* - 1})
\end{eqn}
Since they both satisfy the Ward identity, we obtain,
\begin{eqn}
Q_\Lambda(\Psi - \Psi_*) = 0
\end{eqn}
and using the perturbative expansion for the charge $Q_\Lambda = Q_\Lambda^{(0)} + g Q_\Lambda^{(1)} + \cdots$ we get the constraint
\begin{eqn}
Q_\Lambda^{(0)}\big( e^{\mathcal W_2} (\mathcal W_{n_*} - \mathcal W_{n_*}^*) \big) + O(g) = 0~.
\end{eqn}
Using equations \eqref{QLambda0} and \eqref{Q0eW2=0} we find
\begin{eqn}
&Q_{\Lambda}^{(0)}(e^{ \mathcal W_2} (\mathcal W_{n_*} - \mathcal W_{n_*}^*)) \\
&= -i\Tr e^{\mathcal W_2}\int d^3 k \Lambda_0(\vec k) \frac{\delta (\mathcal W_{n_*} - \mathcal W_{n_*}^*)}{\delta \varphi(\vec k)}
\end{eqn}
which leaves us with the following integral constraint at $O(g^{n_*-2})$,
\begin{eqn}
\Tr \int d^3 k \Lambda_0(\vec k) \frac{\delta (\mathcal W_{n_*} - \mathcal W_{n_*}^*)}{\delta \varphi(\vec k)} = 0~.
\end{eqn}
Since this equation has to be true for arbitrary $\Lambda_0(\vec k)$ it implies that for every $\vec k$ we have
\begin{eqn}
\frac{\delta (\mathcal W_{n_*} - \mathcal W_{n_*}^*)}{\delta \varphi(\vec k)} = 0
\end{eqn}
and since each $\mathcal W_n$ has a fixed dependence on $\varphi$ \eqref{eq:wavefunction_expansion} it implies that the respective wavefunction coefficients must be the same and hence $\mathcal W_{n_*}= \mathcal W_{n_*}^*$. This can be similarly extended for all $n$ and thus proves the uniqueness of the solution. 

We end this section by contrasting our situation with that of full Yang-Mills with standard gauge symmetries \cite{Hatfield:1984dv}. The wavefunction of full YM  is gauge invariant under the standard gauge symmetries (Gauss constraint) which leads to the consistency conditions (soft theorems) for wavefunction coefficients \cite{McFadden:2014nta, Baumann:2020dch, Hui:2018cag, Creminelli:2004yq, Hinterbichler:2013dpa, Hamada:2018vrw}. In general the Gauss constraint restricts the dependence on the longitudinal part of the wavefunction but does not place any constraint on the transverse part. Thus the standard ward identity of the gauge theory does not uniquely fix the wavefunction. This is further confirmed through a careful analysis of the infinite order symmetries leading to consistency conditions for the $n^{th}$ order wavefunction coefficients \cite{Hinterbichler:2013dpa, Berezhiani:2013ewa}. In order to fully determine the wavefunction one needs to additionally impose that it satisfies the Schrodinger equation \cite{Hatfield:1984dv} which gives the perturbative ground state wavefunction. Alternatively solving the perturbiner equations will fully constrain the wavefunction leading to perturbative solutions consistent with the Gauss constraints. However unlike in the present case the ward identities alone do not guarantee a unique solution for the wavefunctions in full Yang-Mills.

\subsection{Alternate Boundary Conditions}\label{ssec:WFbndycond}
We also evaluate the wavefunctions obtained using alternate boundary conditions for the fields and conjugate momentum at $t = 0$. These can be obtained by using appropriate integral transformations from the Dirichlet-Dirichlet wavefunctions as shown below. In self-dual holography with the scalar model, wavefunctions with Robin boundary conditions have been studied in \cite{Skvortsov:2026gtq}.  

Wavefunctions with these alternate boundary conditions are also of interest in the theory with the self-dual gauge fields \cite{Skvortsov:2026dru, Heuveline:2026knw}. Our present discussion does not relate the wavefunctions evaluated with the gauge fields to those with the scalar theory we consider here. However we expect that a particular choice of the parameters with the Robin boundary condition can be mapped to the wavefunctions evaluated with the gauge fields and this is explored in an upcoming paper \cite{upcoming}. 


\subsubsection{Neumann-Neumann}
Wavefunctions with Neumann-Neumann boundary conditions can be obtained from the Dirichlet-Dirichlet wavefunctions by taking a Fourier transform 
\begin{eqn}
\Psi[\pi_\varphi, \pi_{\bar\varphi}] = \int D\varphi D\bar\varphi e^{-i \tr \int_{\Sigma_0} d^3 x (\varphi \pi_\varphi + \bar\varphi \pi_{\bar\varphi})} \Psi[\varphi, \bar\varphi]
\end{eqn}
This is similar to how one obtains the wavefunction in momentum space in quantum mechanics and we review this in \autoref{sec:boundarycond}. As an example we evaluate this upto the two-pt wavefunction coefficient and obtain
\begin{eqn}
\Psi_N[\pi_\varphi, \pi_{\bar\varphi}]= e^{- \int d^3 k \frac{1}{k}\pi_{\varphi}(\vec k) \pi_{\bar\varphi}(- \vec k)} 
\end{eqn}
thus recovering the standard form for the 2-pt wavefunction in this basis.

\subsubsection{Robin}
Robin boundary conditions are implemented through suitable boundary terms. In our case, we hold fixed linear combinations of $\varphi, \pi_{\varphi}, \bar\varphi, \pi_{\bar\varphi}$ on the time slice, following the conventions in \eqref{app:Robin}.

The corresponding wavefunction is obtained from the Dirichlet-Dirichlet
wavefunction through the weighted Fourier transform explicitly given in \eqref{Srobinfinal},
\begin{eqn}
\Psi_{\rm Robin}[\rho,\bar\rho]
&=
\int D \varphi D\bar\varphi e^{i S_\p^{\rm Robin}} \Psi[\varphi,\bar\varphi] .
\end{eqn}
Other examples with some special cases are discussed in \autoref{app:Robin}. 

\subsubsection{$\delta-$Wavefunction}\label{ssec:deltaWF}
There is a special case within this general class of mixed boundary conditions worth pointing out explicitly. This is motivated by the variation of the action in \eqref{scalar_Lag}. The general variation of the action \eqref{scalar_Lag} on the support  of e.o.m is given as,
\begin{eqn}\label{mixedNeumann}
\delta S_{YM} =\Tr \int d^3 x \left( \pi_{\Phi} \delta \Phi - \bar \Phi \delta \pi_{\bar\Phi} \right)
\end{eqn}
Thus without adding any boundary terms the variation is well defined for the choice of Dirichlet-Neumann boundary conditions. However the action in this case becomes proportional to  e.o.m for $\Phi$, and hence vanishes on-shell. This naively leads to a trivial ground state but this is incorrect. The simplest way to see this is by deriving the wavefunction  via a Fourier transform in $\bar\Phi$ from the Dirichlet-Dirichlet wavefunction, 
\begin{eqn}\label{DNfromDD}
\Psi[\Phi, \pi_{\bar\Phi}] = \int D\bar\Phi e^{-i {\scriptsize\Tr}\int_{\Sigma_0} d^3 x (\bar \Phi \pi_{\bar\Phi})} \Psi[\Phi, \bar\Phi]
\end{eqn}
Recalling that the Dirichlet-Dirichlet wavefunction coefficients in the RHS are linearly dependent on  $\bar\Phi(x)$ as given in \eqref{Wonshell} we get
\begin{eqn}\label{deltaPsi}
 \Psi[\Phi, \pi_{\bar\Phi}] &= \int D\bar\Phi e^{i {\scriptsize\Tr}\int_{\Sigma_0} d^3 x \, \bar \Phi (\pi_{\bar\Phi} - \p_t \Phi_{cl}|_{t = 0})} \\
 &\propto \delta(\pi_{\bar\Phi} - \p_t \Phi_{cl}|_{t = 0}) 
\end{eqn}
Thus the resulting wavefunction in this basis  is a delta function on the support of a specific value for the conjugate momentum. This is consistent with its classical equation given in \eqref{conjmomYM}. Note that although the on-shell action does vanish in this case, it does not represent the $\delta-$wavefunction as the saddle point approximation becomes invalid. Hence for these boundary conditions it is more convenient to evaluate the wavefunctions via the Fourier transform.

\section{Soft Theorems: Family by Family  Ward Identities}\label{sec:familysymmsoft}
Since we also have conserved charges corresponding to symmetries family by family it is natural to ask what the implication of their corresponding ward identities are on the wavefunction coefficients. We find that family by family ward identities constrain the {\it soft limits} of wavefunction coefficients.

 For the gauge parameter $\Lambda$ which keeps the state $\Psi = e^{\mathcal W}$ at $t = - \infty$ unchanged the ward identity upto the first family is given as
\begin{eqn}\label{familyWImain}
\frac{1}{\Psi}Q_{\Lambda} \Psi &= \Tr \int d^3 x \Big( - i \tilde \Lambda \frac{\delta \mathcal W}{\delta \varphi} - \bar\varphi \dot{\tilde\Lambda} - i g \bar\varphi [\dot\Lambda, \varphi] \\
&\qquad\qquad + g [\Lambda, \varphi] \frac{\delta \mathcal W}{\delta \varphi} + g[\Lambda, \bar\varphi] \frac{\delta \mathcal W}{\delta \bar\varphi} \Big) = 0
\end{eqn}
We discuss the role of higher families towards the end. We denote the wavefunction coefficient by $\psi_{n+1}(\vec p,T^a;1,\ldots,n)$, where $\vec k_n$ is the momentum of the $\bar\varphi$ leg and $\vec p$ is the momentum of the soft leg. We present the detailed derivation in appendix \ref{app:WIfamily}.

\subsection{$0^{th}$ Family}
Choose the following gauge parameter for the zeroth family
\begin{eqn}
\delta_{\Lambda} \Phi = e^{i \sqrt{2} \a v + i \b w} f^a T^a, \qquad \delta_{\Lambda} \bar\Phi = 0
\end{eqn}
where $\a >0$ and $\b, f^a$ are arbitrary parameters and $T^a$ denote the color matrices. The restriction on the sign of $\a$ is necessitated by the gauge invariance of the BD state. Plugging this into \eqref{familyWImain} and simplifying the expression we get 
\begin{eqn}\label{WNconstraint}
\left.
\sum_{n=3}^{\infty}g^{n-2}
\frac{\delta \mathcal W_n}{\delta\varphi(\vec p)}
\right|_{\vec p=\alpha\widehat z+ \b \widehat w} = 0 , \qquad n \geq 3
\end{eqn}
from which it immediately follows that $\psi_n$ satisfies
\begin{eqn}\label{zerothfamconst}
\Big. \psi_n(\vec p; \{\vec k\}) \Big|_{\vec p = \a \widehat z + \b \widehat w} = \Big. \psi_n(\vec p; \{\vec k\}) \Big|_{p_{\wb} = 0} = 0
\end{eqn}
for $n \geq 3$. 
This is reminiscent of an Adler zero which is standard for shift symmetries \cite{Adler:1964um, Low:2014nga, Bittermann:2022nfh, Du:2024soq}. This completes the analysis for the implication of the Ward identities corresponding to the zeroth family with the details given in \autoref{app:WIfamily}. 

\subsection{$1^{st}$ and Higher Families}\label{sec:higherfam}
We use the colour-ordered coefficients defined in
\eqref{eq:wavefunction_expansion}. 
The zeroth-family constraint \eqref{zerothfamconst} gives a series expansion, wherever the coefficient is regular at $p_{\bar w}=0$,
\begin{equation}
\psi_{n+1}(\vec p,\vec k_1,\ldots,\vec k_n)
=\sum_{m=1}^{\infty}p_{\bar w}^{m}\widehat S_m\psi_n,
\end{equation}
where the equation right now is thought of a formal series expansion around $p_{\bar w} = 0$.
 The first-family Ward identity fixes the linear coefficient exactly at
tree level,
\begin{align}\label{S1soft}
&\widehat S_1\psi_n= -\frac{i}{\sqrt{2}}\delta_{n,2},
\qquad n\geq2.\\
&\quad-\frac{i}{\sqrt{2}\,p_z}
\Big[\psi_n(\vec k_1+\vec p,\ldots,\vec k_n)-\psi_n(\vec k_1,\ldots,\vec k_n+\vec p)
\Big]_{p_{\bar w}=0}
\nno
\end{align}
Only the two legs adjacent to the soft leg in the cyclic ordering appear which in our convention is chosen as 
the first $\varphi$ leg and $\bar\varphi$. In each term, only the
displayed momentum is shifted. 
The first term is the contribution of
$-ig\Tr\int d^3x\,\bar\varphi[\dot\Lambda,\varphi]$
in \eqref{familyWImain} and occurs only in the $3\to2$ point relation. For $n\geq3$, the result is therefore just the difference of the two shifted lower point coefficients. Appendix \ref{app:first_higher_families} derives the color reduction explicitly.

Equation \eqref{S1soft} holds for $p_z>0$ and arbitrary $p_w$. For the higher families, the transformations used here are known only perturbatively in $g$. Their Ward identities therefore constrain the higher coefficients in the series expansion order by order in $g$.  

\subsection{Recovering Diagrammatic Soft Limits}
In a previous paper \cite{Chowdhury:2024dcy} we have evaluated the soft limits of a single class of Witten diagrams whose final result is given as
\begin{eqn}\label{oldsoftop}
\psi_{n+1}\Big|^{\text{class I}}_{\vec p = 0}= \left( \frac{X_{p k_1}}{k_1} \p_{k_1} - \frac{X_{p k_{n}}}{k_{n}} \p_{k_{n-1}} \right) \psi_n
\end{eqn}
This analysis does not capture the soft limit of the full wavefunction coefficient but provides an useful starting point. We show how a special case of soft limits corresponding to the $1^{st}$ family \eqref{S1soft} gives recovers the same result wherever applicable.

Consider the soft factor acting on an $n \geq 3$  point wavefunction $\widehat S_1 \Psi_n$, such that the final term in \eqref{S1soft} does not contribute. To make contact with the soft limits in \cite{Chowdhury:2024dcy} all components of $\vec p \to0$ and therefore it is a special case of the  \eqref{S1soft}. To extract the soft factor in that limit we parametrize the soft momentum as $\vec p = (p_z, 0,0)$ and take the soft limit $p_z \to 0^+$, which ensures that the full vector goes to zero. From \eqref{S1soft} by using $\lim_{\e\to0} \frac{f(x + \e) - f(x)}{\e} = \p_x f$ we immediately get,
\begin{eqn}
\Big.\psi_{n+1} \Big|_{\vec p = 0} = - \frac{i}{\sqrt{2}} p_{\bar w} \big( \p_{k_{1z}} - \p_{k_{n\,z}} \big) \psi_n
\end{eqn}
This result is valid for the full wavefunction coefficient and does not rely on any particular subset of Witten diagrams. For generic $n$ this does not reduce to the soft operator \eqref{oldsoftop} as given in \cite{Chowdhury:2024dcy} as the $z-$directional momentum derivatives on the RHS do not automatically convert to the form in \eqref{oldsoftop}. However for the special case of $4 \to 3$ soft limits they can be shown to agree by using the formula
\begin{eqn}\label{4to3specialid}
\p_{k_{rz}} \psi_3 = - \frac{2k_{ru}}{k_r} \p_{k_r} \psi_3 - \frac{\psi_3}{k_{123}}, \qquad r = 1, 2, 3
\end{eqn}
which follows from chain rule and does not hold at higher points. Thus for this case we get
\begin{eqn}
\psi_{4}  \Big|_{\vec p = 0} = i \sqrt{2} \left( \frac{p_{\bar w} k_{1u}}{k_1} \p_{k_1} - \frac{p_{\bar w} k_{3u}}{k_3} \p_{k_3}\right) \psi_3
\end{eqn}
This reproduces the energy-derivative structure of the diagrammatic soft relation for the $4 \to 3$ limit on the $p_w =0$ branch. At higher multiplicity, converting spatial-momentum derivatives into energy derivatives also produces contributions from the internal momentum dependence. This can be seen from the general identity replacing \eqref{4to3specialid} 
\begin{align}
\partial_{k_{rz}}\psi_n
={}&-\frac{2k_{ru}}{k_r}\,\partial_{k_r}\psi_n
+\left.\partial_{k_r}\psi_n
\right|_{\{k_{au}\},\,\{|\vec k_{\mathrm{int}}|\}}
\nonumber\\
&+\sum_{\substack{\mathrm{int}\\ r\in\mathrm{int}}}
\frac{k_{\mathrm{int},z}}{|\vec k_{\mathrm{int}}|}
\left.\partial_{|\vec k_{\mathrm{int}}|}\psi_n
\right|_{\{k_a\},\,\{k_{au}\}} 
\end{align}
for all legs $a$. Here $\vec k_{\mathrm{int}}=\sum_{a\in\mathrm{int}}\vec k_a$ denotes an internal momentum, and the sum runs over internal channels containing leg $r$. The subscripts on the vertical bars indicate the variables held fixed.

\section{$n$-pt Wavefunction as a Determinant}\label{sec:nptPsi}
In this section we describe how the tree level $n$-pt wavefunction can be obtained recursively from the perturbiner solution and packaged as a determinant of an {\it almost} upper-triangular matrix of order $2^{n-2} -1$, where, the matrix is upper triangular except for its last row.

\subsection{Low point Perturbiners}
We consider the color ordered wavefunction coefficients $\psi_n$ in \eqref{eq:wavefunction_expansion}, with leg $n$ corresponding to $\bar\varphi$ and the factor $g^{n-2}$ extracted as in \eqref{eq:WFexp}. We first introduce the following notations to simplify the discussion. For a consecutive set of numbers $Q$ within $(1, \cdots, n-1)$, labeling the $\varphi$ fields, define,
\begin{eqn}\label{eq:npt_zeta}
\vec k_Q=\sum_{a\in Q}\vec k_a,
\qquad
\zeta_Q \equiv\frac{|\vec k_Q|+k_{Qz}}{k_{Q\bar w}} = \frac{k_{Qw}}{k_{Qu}}
\end{eqn}
where the second equality for $\zeta_Q$ follows from $|\vec k_Q|^2 =k_{Qz}^2 + 2k_{Qw} k_{Q\wb}$ and $k_{Qu} = \frac{|\vec k_Q| - k_{Qz}}{2}$.

The free solution of the bulk e.o.m is $\Phi_0(t,\vec k)=e^{ikt}\varphi(\vec k)$. To obtain the higher point functions it is convenient to introduce the following notations. We denote by $\Phi_{1\cdots m}(t)$ the coefficient of $g^{m-1}e^{i\vec k_{1\cdots m}\cdot\vec x} \varphi_1\cdots\varphi_m$ in the classical solution with $\varphi_a \equiv \varphi(\vec k_a)$. Thus $\Phi_a(t)=e^{ik_at}$, and \eqref{Wonshell} gives
\begin{eqn}\label{PsiPerPhi}
\psi_n(1,\ldots,n)
=\left.i\partial_t\Phi_{1\cdots n-1}(t)\right|_{t=0}.
\end{eqn}
Substituting the plane wave solution into the quadratic term of \eqref{Phi_eom}
gives the first perturbiner correction satisfying,
\begin{eqn}
\left(-\partial_t^2-|\vec k_{12}|^2\right)\Phi_{12}(t)
&=-2i\sqrt{2}X_{12}e^{i(k_1+k_2)t},\\
 \Phi_{12}(0)&=0.
\end{eqn}
where $X_{12} = k_{1u} k_{2\wb} - k_{1\wb} k_{2u}$. Using the Dirichlet Green function given in \eqref{Dgreenfunction}, the solution becomes
\begin{eqn}
\Phi_{12}(t)
=-2\sqrt{2}X_{12}\int_{-\infty}^{0}dt'\,
G_D(t,t',\vec k_{12})\,e^{i(k_1+k_2)t'}.
\end{eqn}
Evaluating the time integral $t'$ results in
\begin{eqn}
\Phi_{12}(t) =  -\frac{2i\sqrt{2}X_{12}}{(k_1 + k_2)^2 - |\vec k_{12}|^2} \Big[ e^{i(k_1 + k_2)t} - e^{i|\vec k_{12}|t} \Big]
\end{eqn}
which satisfies $\Phi_{12}(0)=0$. The denominator can be expressed in terms of $\zeta$ by using the relation $(k_1+k_2)^2-|\vec k_{12}|^2
=2X_{12}(\zeta_2-\zeta_1)$. Using \eqref{PsiPerPhi} we obtain the 3-pt wavefunction coefficient
\begin{eqn}\label{eq:npt_psi3}
\psi_3(1,2,3)
=-i\sqrt{2}\,
\frac{k_3-k_1-k_2}{\zeta_2-\zeta_1}
=\frac{2i\sqrt{2}X_{12}}{k_1+k_2+k_3}.
\end{eqn}

Iterating this process once more results in the 4-pt wavefunction coefficient. As an intermediate step, the solution for the $O(g^2)$ term in the perturbiner is given as,
\begin{align}
\Phi_{123}(t)
={}&-2\int_{-\infty}^{0}dt'\,
G_D(t,t',\vec k_{123})
\\
&\times\varep^{ij}\Big(
\partial_i\Phi_{12}(t')\partial_j\Phi_3(t')
+\partial_i\Phi_1(t')\partial_j\Phi_{23}(t')
\Big).\nno
\end{align}
Note that the derivatives act on the full modes $e^{i\vec k_Q\cdot\vec x}\Phi_Q(t)$ even though we have suppressed the spatial plane wave dependence for brevity. Evaluating the time integrals and simplifying using the relations
between $X_{ij}$ and $\zeta$ gives
\begin{align}\label{eq:npt_cubic}
\Phi_{123}(t)
={}&-\frac{2}
{(\zeta_2-\zeta_1)(\zeta_3-\zeta_2)}
\left[
e^{i(k_1+k_2+k_3)t}-e^{i|\vec k_{123}|t}
\right]
\\
&+i\sqrt{2}\left[
\frac{\Phi_{1,23}(t)}{\zeta_3-\zeta_2}
+\frac{\Phi_{12,3}(t)}{\zeta_2-\zeta_1}
\right].\nno
\end{align}
This form is analogous to the Berends-Giele recursion relations \cite{Berends:1987me, Mizera:2018jbh, LipinskiJusinskas:2026ctz,Rosly:1996vr,Rosly:1997ap}. 
Each grouped label in $\Phi_{1,23}, \Phi_{12,3}$ is treated as a single leg with energy $|\vec k_{23}|$ or $|\vec k_{12}|$, respectively. Applying \eqref{PsiPerPhi} and using $|\vec k_{123}|=k_4$ gives the wavefunction coefficients,
\begin{align}\label{eq:npt_psi4}
\psi_4
={}&-2\,
\frac{k_4-k_1-k_2-k_3}
{(\zeta_2-\zeta_1)(\zeta_3-\zeta_2)}\\
&+i\sqrt{2}\left[
\frac{\psi_3(1,23,4)}{\zeta_3-\zeta_2}
+\frac{\psi_3(12,3,4)}{\zeta_2-\zeta_1}
\right].\nno
\end{align}
Which can be checked to agree with the explicit evaluation using the Witten diagram method given in \eqref{4ptwittendirichlet}. In the upcoming section we demonstrate how this can be understood via a determinant structure.

\subsection{Determinant Structure}\label{ssec:detstructure}
The boundary wavefunction can be organised into a determinant structure. This packages the recursive structure of the perturbiner expansion. 
We provide the general form of the result and demonstrate how it works in the simplest examples upto 5-pt. Further details  will be explained in an upcoming paper \cite{upcoming}.

The final form of $\psi_n$ can be expressed as a determinant of a $M\equiv 2^{n-2} - 1$ dimensional matrix $(A_{n})_{ij}$ which is given as
\begin{eqn}\label{eq:npt_determinant}
\psi_n(1,\ldots,n)
=(-i\sqrt{2})^{n-2}
k_{n\bar w}\det A_{n},
\quad n\geq4.
\end{eqn}
The matrix $(A_n)_{ij}$ is constructed as follows. First, define the column partition as
$\tau_j=(Q_1|\cdots|Q_s)$. For $i<M$, set $(A_n)_{ij}=0$ unless $\tau_j$ {\it refines} $\tau_i$, meaning that every cut of $\tau_i$ remains
present in $\tau_j$. For example if $\tau_i = (1|234)$ and $\tau_j = (1|23|4)$ then $\tau_j$ refines $\tau_i$. For the case when $\tau_j$ refines $\tau_i$ we have,
\begin{eqn}\label{eq:npt_refinement}
(A_{n})_{ij}
&=\prod_r'\frac{1}{\zeta_{Q_{r+1}}-\zeta_{Q_r}},
\qquad i<M,\\
(A_n)_{Mj}
&=(\zeta_{Q_s}-\zeta_P)
\prod_{r=1}^{s-1}
\frac{1}{\zeta_{Q_{r+1}}-\zeta_{Q_r}}
\end{eqn}
The rows and columns are indexed by all consecutive partitions of $P=(1,\ldots,n-1)$ into at least two nonempty blocks, ordered by increasing number of blocks, with the fully split partition last. In the first line the primed product runs over adjacent blocks $Q_r,Q_{r+1}$ contained in the same block of $\tau_i$. An empty product equals one, so $(A_n)_{ii}=1$ for $i<M$. This gives the almost upper-triangular structure for the matrix. 

For the $n =3$ case the answer is not directly expressed as a single determinant. For the 3-pt case there is only a single partition allowed for the $\varphi$ legs $1|2$, which results in a one-dimensional matrix
$A_{3}=[(\zeta_2-\zeta_{12})/(\zeta_2-\zeta_1)]$ and the corresponding wavefunction coefficient is 
\begin{eqn}\label{psi3detA3}
\psi_3 = -i \sqrt{2} (k_{1\bar w} + k_{3\bar w} \det A_{3})
\end{eqn}
Using the definition of $\zeta_a$ given in \eqref{eq:npt_zeta} and the identity 
\begin{eqn}\label{Xzetaidentity}
X_{AB}(\zeta_A - \zeta_B) = - |\vec k_A| |\vec k_B| + \vec k_A \cdot \vec k_B
\end{eqn}
where $A,B$ are the two blocks of momenta, it can be verified that this matches the perturbiner calculation \eqref{eq:npt_psi3} and the Witten diagram computation given in \eqref{3ptwitten} (upto normalization). The RHS of the identity \eqref{Xzetaidentity} resembles the product of four momentum between the two blocks of momenta $A, B$. This identity results in the simplified forms of expressions.

At four points, there are three possible ordered partitions. These are given as
$\tau_1 = 1|23$, $\tau_2 = 12|3$, and $\tau_3 = 1|2|3$. Thus $\tau_3$ refines $\tau_1, \tau_2$. These three partitions label the three terms in \eqref{eq:npt_psi4}. Using \eqref{eq:npt_refinement} the matrix $A_4$ can be read off as
\begin{eqn}\label{eq:npt_A123}
\setlength{\arraycolsep}{3pt}
A_{4}=
\begin{pmatrix}
1 & 0 & \frac{1}{\zeta_3-\zeta_2}\\[3pt]
0 & 1 & \frac{1}{\zeta_2-\zeta_1}\\[3pt]
\frac{\zeta_{23}-\zeta_{123}}{\zeta_{23}-\zeta_1}
&
\frac{\zeta_3-\zeta_{123}}{\zeta_3-\zeta_{12}}
&
\frac{\zeta_3-\zeta_{123}}
{(\zeta_2-\zeta_1)(\zeta_3-\zeta_2)}
\end{pmatrix}.
\end{eqn}
Using \eqref{eq:npt_determinant} we  obtain $\psi_4 = - 2 k_{4\bar w} \det A_{4}$ and expanding the determinant along the final column shows the recursive structure given in \eqref{eq:npt_psi4} which 
in turn agrees with the Witten diagram evaluation (upon adding both channels) given in \eqref{4ptwittendirichlet}. As another simple check it can be verified that this satisfies the soft limit corresponding to the zeroth family. When $p_{r\bar w} \to 0$, for $r \in (1, \cdots, n-1)$, we have $\zeta_{r} \to \infty$, which implies that the matrix develops either two proportional columns or a zero column, making the vanishing of its determinant manifest. The same mechanism also extends to higher points as shown in \ref{app:fivept}. Note that the wavefunction coefficient does not contain the total energy pole and hence vanishes in the flat space limit as expected. 

\subsection{Determinant Representation for Flat-Space SDG}\label{app:SDGdet}
The structures above generalize to SDG. We provide the complete details in an upcoming paper and only display an example here \cite{upcoming}. The boundary on-shell action for SDG is given as
\begin{eqn}\label{SDGbndy}
S_{\partial D}^{SDG}
=\int_{\Sigma_0}d^3x\,
\bar\varphi
\left(1+\frac{\kappa}{2}\partial_{\bar w}^{\,2}\varphi\right)
\partial_t\phi\big|_{t=0}.
\end{eqn}
This perturbiner solution for the classical field  admits a similar recursive solution as the SDYM case which can again be reorganized into a determinant. For the 4-pt wavefunction coefficient this is given as
\begin{eqn}\label{eq:SDG_A4}
\psi_4^{\mathrm{SDG}}=\det 
\begin{pmatrix}
1&0&0&\frac{X_{23}}{\zeta_3-\zeta_2}\\[6pt]
0&1&0&\frac{X_{13}}{\zeta_3-\zeta_1}\\[6pt]
0&0&1&\frac{X_{12}}{\zeta_2-\zeta_1}\\[6pt]
q_1&q_2&q_3&q_4
\end{pmatrix}
\end{eqn}
where the entries in the last lows are 
\begin{eqn*}
q_a=
\left(k_4-k_a-|\vec k_{bc}|\right)
\frac{X_{a,bc}}{\zeta_{bc}-\zeta_a}
+\frac12\left(
k_{a\bar w}^{\,2}|\vec k_{bc}|
+k_{bc\bar w}^{\,2}k_a
\right),
\end{eqn*}
for $(a,b,c)=(1,2,3),(2,3,1),(3,1,2)$, and 
\begin{align*}
q_4=
\sum_{\mathrm{cyc}(a,b,c)}
\frac{X_{bc}}{\zeta_c-\zeta_b}
&\Bigg[
(k_4-k_1-k_2-k_3)
\frac{X_{ab}}{\zeta_b-\zeta_a}\\
&\quad+\frac12\left(
k_{a\bar w}^{\,2}(k_b+k_c)
+k_{bc\bar w}^{\,2}k_a
\right)
\Bigg].\nno
\end{align*}
There are numerous examples in the literature where determinant structures appear in the expressions for the scattering amplitude \cite{Hodges:2012ym, Feng:2012sy, Cachazo:2013hca, Cachazo:2012kg, Cachazo:2012da}. The most notable example is of the Hodge's formula for computing the gravitational MHV amplitudes \cite{Hodges:2012ym} where the result can be directly written as a reduced determinant constructed in kinematic space. This provides a remarkably compact expression compared to diagrammatic or recursive expansions. Our representation does not yet establish a comparable computational simplification as the matrix dimension grows exponentially with the number of particles, whereas the matrix in Hodges' formula grows linearly. Nevertheless, it organizes the recursive boundary corrections into a single determinant. It would be interesting to determine whether further use of the symmetries or a more suitable choice of variables leads to a more compact representation.

\section{Conclusion}\label{sec:conc}
In this paper we studied symmetries and the ground state wavefunctions for self-dual Yang-Mills and Gravity. We worked with the scalar actions for the two theories which admit an infinite tower of symmetries. We resummed these and demonstrated that they reduce to single symmetries with a general harmonic parameter. 

We constructed the Noether charges associated to these resummed symmetries. The corresponding Ward identities turn out to be powerful enough to fully fix the tree-level ground state wavefunction for both theories. This is in contrast with the standard Ward identities of Yang-Mills and Gravity which only fix a part of the answer. 

The wavefunction coefficients are expected to satisfy consistency conditions in the form of soft limits. We show how these arise as a consequence of the Ward identities corresponding to the tower of the symmetries. For example, the zeroth level in the tower gives $\psi_n(\vec p)\Big|_{p_{\bar w} = 0} = 0$, with the higher levels constraining the expansion in $p_{\bar w}$ where $p$ labels the momentum of any of the positive helicity legs. This allows us to recover the previously obtained diagrammatic soft limits \cite{Chowdhury:2024dcy} as a special case. 

As a curiosity we find the $n$-pt wavefunction coefficient can be expressed as a single determinant in terms of the momenta of external particles. While this has a flavour of Hodge's formula for gravitational MHV amplitudes, the matrix is much larger. This passes the standard consistency checks including the soft limits described above.

We are currently working to understand how our results fit within full Yang-Mills and Gravity theories which requires a careful treatment term of the boundary terms \cite{upcoming}. To this end we have provided the rules for mapping our wavefunction to a generic wavefunction corresponding to a class of Robin boundary conditions.

\section*{Acknowledgments}
We are grateful to Song He, Arthur Lipstein, and Yuyu Mo for useful discussions. 

SN is supported by STFC consolidated grant UKRI4226. CC is supported by the National Natural Science Foundation of China and by the New Cornerstone Science Foundation.

Use of AI: All ideas and the logical steps were developed by the authors with the exception of the determinant interpretation for the $n$-pt case. We have used the publicly available model for ChatGPT Astra. The authors have independently checked all the results generated by AI and take full responsibility of the paper.

\bibliography{references}

\appendix
\onecolumngrid

\section{Symmetry Calculations}
\label{Symmetry_calc_app}
\subsection{Symmetry parameter}
\label{App:Symmetry parameter}
The summed parameters appearing when writing the sum
\be 
\d\Phi\equiv\sum_{n=0}^{\infty}\d_{\L_n}\Phi
\ee 
are
\be \label{defLtilde}
\tilde{\L}=\L_0+\sum_{n=1}^{\infty}\frac{1}{n}\left(x^i\varepsilon_{ij}\partial^j\right)\Lambda_n
=\sum_{n=0}^{\infty}\frac{1}{n!}\left(x^i\varepsilon_{ij}\partial^j\right)^n\bar{\L}_n(v,w)
\ee 
and 
\be \label{defL}
\L=\sum_{n=1}^{\infty}\Lambda_n
=\sum_{n=1}^{\infty}\frac{1}{(n-1)!}\left(x^i\varepsilon_{ij}\partial^j\right)^{n-1}\bar\Lambda_n(v,w)
=\sum_{n=0}^{\infty}\frac{1}{n!}\left(x^i\varepsilon_{ij}\partial^j\right)^{n}\bar\Lambda_{n+1}(v,w)
\ee 
In the above we used \eqref{0th} and \eqref{LAmbda_n_def}. From the definitions above, it is immediate to see that \eqref{LtildeL} is satisfied. Now we will show that $\L$ and $\tilde\L$ are general harmonic functions.

Consider a function $f=f(u,v,w,\wb)$. Its Taylor expansion in the $u,\wb$ coordinates is
\be \label{Taylor}
f(u,v,w,\wb)=\sum_{p,q\geq 0}\frac{u^p\wb^q}{p!q!}a_{p,q}(v,w)
\ee 
Now assume 
\be 
\square f=0 \, .
\ee 
This imposes the following condition on the $a$'s:
\be 
\partial_v a_{p+1,q}=\partial_w a_{p,q+1} \, .
\ee 
It will be useful to rewrite the above in terms of the total degree $n\equiv p+q+1$, with $p\geq0$ and $q\geq0$:
\be \label{constr}
\partial_v a_{p+1,n-p-1}=\partial_w a_{p,n-p} 
\ee 
The above means that we can write the coefficients $a$ in terms of a potential $\bar{a}$ as
\be \label{A_potential}
a_{l,n-l}(v,w)=\p_w^l\p_v^{n-l}\bar{a}_n(v,w),\quad 0\leq l\leq n
\ee
\begin{proof}
We can prove the above by recursion in $n$. Firstly, $a_{0,0}(v,w)\equiv \bar a_0$ is unconstrained by \eqref{constr}. At $n=1$, we have 2 parameters
\be 
a_{1,0},\quad a_{0,1}
\ee 
constrained by one equation
\be 
\p_v a_{1,0}=\p_w a_{0,1}
\ee 
To solve this first observe 
\be 
a_{1,0}=\int_{v_0}^v \p_w a_{0,1}(\tilde{v},w)d\tilde{v}+f(w) 
\ee 
Since $f(w)$ is an arbitrary function, we can introduce the notation $f(w)=\partial_w g(w)$, and denote
\be 
\bar{a}_1=\int_{v_0}^v  a_{0,1}(\tilde{v},w)d\tilde{v}+g(w)
\ee 
with solution
\be 
a_{1,0}=\p_w\bar{a}_1,\quad a_{0,1}=\p_v \bar{a}_1 \, ,
\ee
as required. Now, let us prove the general case \eqref{A_potential} by induction.

Now consider a set of $n+2$ functions of $(v,w)$, which we denote by
\be 
F_0,\quad F_1,\quad...\quad F_{n},\quad F_{n+1} \, ,
\ee 
They satisfy $n+1$ constraints
\be \label{App:n_constraints}
\p_v F_0=\p_w F_1,\quad\p_v F_1=\p_w F_2,\quad...\quad
\p_v F_{n-1}=\p_w F_{n},\quad \p_v F_{n}=\p_w F_{n+1}
\ee  
Assume that we can solve the first $n$ of the constraints above to get
\be 
F_l=\p_w^{n-l}\p_v^{l}\bar{F}_{n}\, ,\quad
0\leq k\leq n
\ee 
for some potential $\bar{F}_{n}$. We verified the above explicitly for the case $n=1$, with $F_0=a_{1,0}$ and $F_1=a_{0,1}$. In particular, we have 
\be 
\label{App:F_n_min_one}
F_{n}=\p_v^{n}\bar{F}_{n} \, .
\ee 
From the final equation in \eqref{App:n_constraints}, we get
\be \label{last_n_eq_sol}
F_{n}=\p_w\bar{F},\quad
F_{n+1}=\p_v\bar{F}
\ee
for some potential $\bar{F}$.  Then from \eqref{App:F_n_min_one} and \eqref{last_n_eq_sol}, we get
\be 
\p_v^{n}\bar{F}_{n}=\p_w\bar{F}
\ee
which can be solved to give (As in the $n = 1$ case, possible integration functions are absorbed into the definition of $\bar{F}_{n+1}$.)
\be 
\bar{F}_{n}=\partial_w \bar{F}_{n+1},\quad
\bar{F}=\p_v^{n}\bar{F}_{n+1} \, ,
\ee 
where we have introduced the potential $\bar{F}_{n+1}$. Finally, this allows us to write 
\be 
F_l=\p_w^{n+1-l}\p_v^{l}\bar{F}_{n+1}\, ,\quad
0\leq k\leq n+1
\ee 
and then with the identification 
\be 
a_{l,n+1-l}=F_{n+1-l}
\ee
and relabeling $n+1\to n$ we have proven \eqref{A_potential}.
\end{proof}
Then, returning to $f$ in \eqref{Taylor} and rearranging the Taylor expansion
\be 
\begin{aligned}
f(u,v,w,\wb)&=\sum_{n\geq 0}\sum_{p=0}^n
\frac{u^p\wb^{n-p}}{p!(n-p)!}a_{p,n-p}(v,w)\\
&=\sum_{n\geq 0}\sum_{p=0}^n
\frac{u^p\wb^{n-p}}{p!(n-p)!}\p_w^p \p_v^{n-p}\bar{a}_n(v,w)\\
&=\sum_{n\geq 0}\frac{\left(u\p_w+\wb\p_v\right)^n}{n!}\bar{a}_n(v,w)\\
&=\sum_{n\geq 0}\frac{\left(x^i\varepsilon_{ij}\partial^j\right)^n}{n!}\bar{a}_n(v,w)
\end{aligned}
\ee 
 Finally, with the identification 
$\bar{a}_{n}=\bar\L_n,\ n\geq 0$
 we recover $\tilde\L$ as given in \eqref{defLtilde} and for 
$\bar a_n=\bar\L_{n+1},\ n\geq 0$ we get $\L$ as given in \eqref{defL}. The proof above is to be understood locally, at the level of a formal power series in the variables.

\subsection{Symmetry transformation to all orders}\label{app:allorders}
We can extend the transformations in \eqref{resummed_symm} and \eqref{d_bar_phi_o_1} formally to all orders in $g$. We emphasize that the exact form of the symmetry transformation is highly non-unique, as $\Phi$ and $\bar\Phi$ can vary independently. We leave for future work a more detailed study of the best choice for the symmetry, as well as its algebra. For the purposes of this article, it is sufficient to find a symmetry preserving the e.o.m. and the action. As we show in \autoref{sec:pert} it is the generality of the (harmonic) parameter that ensures the exactness of the Ward identity (though it is possible certain choices could simplify calculations in practice).  

First, let

\be 
\delta_\L\Phi=\sum_{n=0}^{\infty}g^{n}\delta_\L^{(n)}\Phi
\ee 
with $\delta_\L^{(0)}\Phi$ and $\delta_\L^{(1)}\Phi$ as given in \eqref{resummed_symm}.
The higher orders are given by
\be\label{nthordervar}
\delta_\L^{(n)}\Phi=\left(-\frac{2i}{\square}  \varepsilon^{ij} [\p_i \Phi, \p_j \cdot]\right)^{n-2}\delta_\L^{(2)}\Phi\equiv F_n[\Phi]\, , \quad 
n\geq 2
\ee 
where
\be 
\delta_\L^{(2)}\Phi=-\frac{1}{\square}\varepsilon^{ij}\left(
[\L,[\partial_i\Phi,\partial_j\Phi]]+2i[\p_i\d^{(1)}\Phi,\p_j\Phi]
\right)
\ee
For practical calculations, one must a choose consistent integral representation for the $\frac{1}{\square}$ operator (see e.g. below \eqref{perteq}).

The transformation of $\bar{\Phi}$ at order $1$ is given in \eqref{d_bar_phi_o_1} and at higher orders it reads
\be 
\delta_\L^{(n)}\bar\Phi=\frac{2i}{\square} \left[\partial^i \left(-\frac{2i}{\square}  \varepsilon^{ij} [\p_i \Phi, \p_j \cdot] \right)^{n-1}\bar{\Phi}_{\mathfrak{h}},\partial_i\L\right]
+\sum_{k=2}^{n}F'_k\left(-\frac{2i}{\square}  \varepsilon^{ij} [\p_i \Phi, \p_j \cdot]\right)^{n-k}\bar{\Phi}_{\mathfrak{h}}\, , \quad 
n\geq 2
\ee 
where
\be 
F'_k[s]\equiv \frac{d}{d\varepsilon}F_k[\Phi+\varepsilon s]\Big|_{\varepsilon=0}
\ee 
These symmetries preserve the e.o.m. and the action in \eqref{scalar_Lag} varies up to a boundary term which vanishes on-shell, thus not contributing to the charge. The additional Dirichlet-Dirichlet boundary term in \eqref{SYMDirBndy} will give a contribution to the charge. 

The gravity transformations to all orders are immediately obtained via the double copy replacements in \eqref{DCrules}.

\section{Further Details of the Wavefunctions} \label{app:Witten}
We give further details of the wavefunction and the ward identities in this appendix.
\subsection{Witten Diagrams}
In this appendix we review the computation of the wavefunction coefficients using Witten diagrams, perturbiner method and explain the origin of the integral transforms used to go from one basis to another. For pedagogy we focus on the SDYM case as the SDG case is analogous. 

\subsubsection{Dirichlet}
The propagators are given as
\begin{eqn}\label{Dgreenfunction}
K_D^p(t) =  e^{i p t}, \qquad G_D(t, t', \vec y) = \frac{1}{2y} \Big[ e^{-i y |t - t'|} - e^{i y (t + t')} \Big]
\end{eqn}
Note that the bulk-bulk propagator propagates between the fields $\bar\varphi$ and $\varphi$ whereas the bulk-boundary propagators are normalized to such that they propagate between the same field. Diagrammatically we denote these as $- \leftrightarrow \varphi$ and $+ \leftrightarrow \bar\varphi$. The vertex factor connecting three bulk points is shown as
\begin{eqn}
V(P, Q) = \begin{tikzpicture}[baseline]
\draw[thick] (-1.75, 1) -- (1.75,1);
\draw (-1, 0.25) -- (0,-1);
\draw (0, 0.25) -- (0,-1);
\draw (1, 0.25) -- (0,-1);
\node at (-1,0.5) {\scriptsize P};
\node at (0,0.5) {\scriptsize Q};
\node at (1,0.5) {\scriptsize R};
\node at (-0.25, -1) {\small +};
\node at (-0.15, -0.65) {\small +};
\node at (0.15, -1) {\small -};
\node at (-1.15, 0.25) {\small -};
\node at (-0.15, 0.25) {\small -};
\node at (1.15, 0.25) {\small +};
\end{tikzpicture}  = \p_u P \p_{\bar w} Q - \p_{\bar w} P \p_u Q
\end{eqn}
The 3-pt function is determined by the following Witten diagram (upto unimportant normalization factors)
\begin{eqn}\label{3ptwitten}
\psi_3(k_1, k_2, k_3)= \begin{tikzpicture}[baseline]
\draw[thick] (-1.75, 1) -- (1.75,1);
\draw (-1, 1) -- (0,-1);
\draw (0, 1) -- (0,-1);
\draw (1, 1) -- (0,-1);
\node at (-1,1.25) {$1^-$};
\node at (0,1.25) {$2^-$};
\node at (1,1.25) {$3^+$};
\node at (-0.25, -1) {\small -};
\node at (-0.15, -0.5) {\small -};
\node at (0.15, -1) {\small +};
\end{tikzpicture} &= \intnsinf dt K_D^{k_1}(t) K_D^{k_2}(t) K_D^{k_3}(t) 
= X_{12} \intnsinf dt e^{i k_{123} t} = \frac{X_{12}}{k_{123}}
\end{eqn}
where $X_{12} = k_{1u} k_{2\bar w} - k_{1\wb} k_{2u}$. Similarly for the 4-pt function, in the $s-$channel we have,
\begin{align}\label{4ptwittendirichlet}
\psi_4^{(s)}(k_1, k_2, k_3, k_4) = \begin{tikzpicture}[baseline]
\draw[thick] (-1.75, 1) -- (1.75,1);
\draw (-1.5, 1) -- (-1.25,-0.5);
\draw (-1, 1) -- (-1.25,-0.5);
\draw (1.5, 1) -- (1.25,-0.5);
\draw (-1.25,-0.5) -- (1.25,-0.5);
\draw (1, 1) -- (1.25,-0.5);
\node at (-1.5,1.25) {$1^-$};
\node at (-1,1.25) {$2^-$};
\node at (1,1.25) {$3^-$};
\node at (1.5,1.25) {$4^+$};
\node at (-1.45, -0.5) {\small -};
\node at (-1.1, -0.35) {\small -};
\node at (-1.15, -0.65) {\small +};
\node at (1.45, -0.5) {\small +};
\node at (1.1, -0.35) {\small -};
\node at (1.15, -0.65) {\small -};
\end{tikzpicture} &= V(1,2) V(3, \vec y) \intnsinf dt_1 dt_2 \, K_D^{k_1}(t_1) K_D^{k_2}(t_1) K_D^{k_3}(t_2) K_D^{k_4}(t_2) G_D(t_1, t_2, y) \nno \\
&= \frac{X_{12} X_{34}}{(k_{12} + y)(k_{34} + y)(k_{12} + k_{34})}
\end{align}
These solutions can also be obtained by solving the perturbiner equation. 

\subsubsection{Neumann}
The bulk-boundary propagator differs by an overall factor whereas bulk-bulk propagator in this case differs by the boundary term
\begin{eqn}
K_N^{p}(t) = \frac{-i}{p} e^{i p t}, \qquad G_N(t, t', \vec y) = \frac{1}{2y} \Big[ e^{-i y |t - t'|} + e^{i y (t + t')} \Big]
\end{eqn}
By repeating the same exercise as the Dirichlet case and performing the time integrals we obtain the following 3 and 4-pt functions,
\begin{eqn}\label{3-4ptNeumann}
\psi_3^N(k_1, k_2, k_3) &=  \frac{i}{k_1 k_2 k_3} \frac{X_{12}}{k_{123}} \\
\psi_4^N(k_1, k_2, k_3, k_4) &= \frac{1}{k_1 k_2 k_3 k_4} \frac{X_{12} X_{34}}{(k_{12} + y)(k_{34} + y)} \left( \frac{1}{k_{12} + k_{34}} + \frac{1}{y} \right)
\end{eqn}
The reader might note the similarity of these expressions with the ones obtained by computing the in-in correlators. Indeed this is evident from the description of the wavefunction with Neumann boundary conditions as a Fourier transform of the wavefunction with Dirichlet boundary condition as shown in section \ref{app:neumannbc}.

\subsection{Wavefunctions for different Boundary conditions via Fourier Transform}\label{sec:boundarycond}
We review how the wavefunctions with alternate boundary conditions are obtained from the ones evaluated with Dirichlet boundary condition. It is useful to illustrate this in a simple quantum mechanical example where the analog of the Dirichlet boundary condition is the position basis. In this case the most notable example is that of the coherent state. 

Consider the operator $\widehat\rho_\a = \widehat p + \a \widehat x$ where $\a$ is a $c$-number. The representation of a state $\ket\psi$ in the $\rho$ basis can be obtained in terms of its representation in the $x$-basis, $\braket{x|\psi}$, by introducing the complete set of states,
\begin{eqn}\label{completion}
\braket{\rho_\a|\Psi} =\int dx \braket{\rho_\a|x} \braket{x|\psi} = \int dx \braket{\rho_\a|x} \Psi(x)
\end{eqn}
The overlap $\braket{\rho_\a|x}$ can be determined from the eigenvalue equation for $\ket{\rho_\a}$,
\begin{eqn}
\widehat \rho_\a \ket{\rho_\a} = \rho_\a \ket{\rho_\a}
\end{eqn}
upon contracting both sides with $\bra{x}$ we get $\braket{x|\widehat \rho_\a| \rho_\a} = \rho_\a \braket{x|\rho_\a}$. This gives the following differential equation
\begin{eqn}
(-i \p_x + \a x)\braket{x|\rho_\a} = \rho_\a \braket{x|\rho_\a}
\end{eqn}
for which we obtain a well known solution $\braket{x|\rho_\a} = e^{i (\rho x - \frac{\a x^2}{2})}$. Plugging this in \eqref{completion} we obtain 
\begin{eqn}
\braket{\rho_\a|\psi} = \int dx e^{i (\rho x - \frac{\a x^2}{2})} \psi(x)
\end{eqn}
This gives the the wavefunction for generic boundary conditions (for further details see \cite{Papadimitriou:2007sj}).

\subsubsection{Relation with In-In Correlators}\label{app:neumannbc}
Specializing to the case of Neumann boundary condition for fields one obtains 
\begin{eqn}
\tilde \Psi[\pi] = \int D\phi \, e^{-i \int \pi \phi} \Psi[\phi]
\end{eqn}
By expanding $\Psi[\phi]$ in the fields and retaining terms till second order in the coupling we get
\begin{eqn}
\tilde\Psi[\pi] = \int D\phi e^{-i \pi \phi} e^{-\int_{12} \psi_2 \phi \phi} \Big[ 1+ \int_{123} \psi_3 \phi \phi \phi + \int_{1234} \psi_4 \phi \phi \phi \phi + \frac{1}{2}  \left( \int_{123} \psi_3 \phi \phi \phi \right)^2 + \cdots \Big]
\end{eqn}
where we have have suppressed the momentum dependence for brevity. The integral can be easily done using standard Gaussian integrals. By collecting terms of $O(\pi^4)$ in $\log\tilde\Psi[\pi]$ we obtain the expressions in \eqref{3-4ptNeumann}. For real tree level wavefunctions this is proportional to the in-in correlator. 

\subsection{Robin-Robin boundary conditions}\label{app:Robin}
Consider the general linear combinations
\be 
\begin{aligned}
R=&a\pi_\Phi +b\pi_{\bar{\Phi}}+i\left(c\Phi+d\bar{\Phi} \right) \\
\bar R=&e\pi_\Phi +f\pi_{\bar{\Phi}}+i\left(h\Phi+j\bar{\Phi} \right) 
\end{aligned}
\ee 
where $a,b,c,d,e,f,h,j$ are arbitrary complex constants. We want these to commute under the canonical Poisson bracket. Using 
\be 
\{\Phi(\vec{x}),\pi_\Phi(\vec{y})\}_{\text{can}}
=\{\bar\Phi(\vec{x}),\pi_{\bar\Phi}(\vec{y})\}_{\text{can}}=\delta^3(\vec{x}-\vec{y})
\ee 
with all others vanishing. Requiring
\be \label{canonical_constrant_constants}
\{R(\vec{x}),\bar R(\vec{y})\}_{\text{can}}=0
\ee
we get
\be 
i(ce+df-ah-bj)\delta^3(\vec{x}-\vec{y})\quad\Rightarrow\quad  ce+df=ah+bj
\ee 
Now assume that 
\be 
\Delta\equiv af-be\neq 0 
\ee 
i.e. the momentum coefficient matrix is invertible. Note that this excludes Robin-Dirichlet and Dirichlet-Robin boundary conditions, which are treated separately in \autoref{App:Robin-Dirichlet} and \autoref{App:Dirichlet-Robin}, respectively. We can define
\be 
\bar\rho=\frac{eR-a\bar{R}}{\Delta}=-\pi_{\bar{\Phi}}-i\left(\frac{ah-ec}{\Delta}\Phi +\frac{aj-ed}{\Delta}\bar{\Phi} \right)
,\quad \rho=\frac{-fR+b\bar{R}}{\Delta}=-\pi_\Phi-i\left(\frac{fc-bh}{\Delta}\Phi +\frac{fd-bj}{\Delta}\bar{\Phi} \right)
\ee 
so that
\be 
\delta R=\delta\bar{R}=0\quad\Leftrightarrow\quad \delta\rho=\d\bar\rho=0
\ee 
Finally, using \eqref{canonical_constrant_constants}, we can relabel 
\be 
\alpha=\frac{fc-bh}{\Delta},\quad \beta=\frac{aj-ed}{\Delta},\quad 
\kappa=\frac{ah-ec}{\Delta}=\frac{fd-bj}{\Delta}
\ee 
to get 
\be 
\begin{aligned}
\bar\rho=&-\pi_{\bar{\Phi}}-i\left(\kappa\Phi+\beta\bar\Phi \right)\\
\rho=&-\pi_{\Phi}-i\left(\alpha\Phi+\kappa\bar\Phi \right)\\
\end{aligned}~.
\ee 
The appropriate boundary action to add to $S_{YM} + S_{YM}^{\p D}$ at $t = 0$ is given as
\begin{eqn}\label{Srobinfinal}
S_{\p}^{\rm RR} = \int_{\Sigma_0} d^3 x \Tr \Big[ \Phi \rho + \bar\Phi \bar\rho + \frac{i\a}{2}\Phi^2 + i \kappa \bar\Phi \Phi + \frac{i\b}{2} \bar\Phi^2 \Big]
\end{eqn}
through which it can be verified that variation of $S_D + S_\p^{\rm Robin} $ vanishes on the support of the e.o.m at $t = 0$ since, 
\begin{eqn}
\delta(S_D + S_{\p}^{\rm RR}) = \int_{\Sigma_0} d^3x \Tr(\Phi \delta\rho + \bar\Phi \delta\bar\rho)
\end{eqn}
which is true for arbitrary variations. 

\subsubsection{Robin-Dirichlet}
\label{App:Robin-Dirichlet}
The most general definition compatible with \eqref{canonical_constrant_constants} is 
\be 
\begin{aligned}
R&=\pi_{\Phi}+i\left(c\Phi+d\bar{\Phi}\right)\, \qquad \bar{R}=\bar{\Phi}
\end{aligned}
\ee 
The appropriate boundary action to add to $S_{YM} + S_{YM}^{\p D}$ at $t = 0$ is now 
\be 
S_\partial^{RD}=\int_{\Sigma_0}d^3x \Tr\left[-\Phi R+\frac{ic}{2}\Phi^2 +id\Phi\bar\Phi\right]
\ee

\subsubsection{Dirichlet-Robin}
\label{App:Dirichlet-Robin}
Here, again requiring compatibility with \eqref{canonical_constrant_constants} gives 
\be 
\begin{aligned}
R&=\Phi,\qquad
\bar R=\pi_{\bar\Phi}+i\left(h\Phi+l\bar\Phi \right)
\end{aligned}
\ee 
with boundary term
\be 
S_\partial^{DR}=\int_{\Sigma_0}d^3x \Tr\left[-\bar\Phi \bar R+\frac{il}{2}\bar\Phi^2+ih\Phi\bar\Phi \right]
\ee

\subsection{Perturbiner Solution: 5-pts}\label{app:fivept}
We show how the determinant formula for the 5-pt example matches with the perturbiner solution. We first obtain the five-point coefficient from the perturbiner. For a consecutive block
$Q=(a,\ldots,b)$, the regular bulk coefficient is
\begin{eqn}\label{eq:app_bulk_coeff}
\Phi_Q^{\rm bulk}(t)
=
\frac{(-i\sqrt{2})^{b-a}
e^{it\sum_{\ell=a}^{b}k_\ell}}
{\prod_{\ell=a}^{b-1}\Delta_{\ell,\ell+1}},
\end{eqn}
where $\Delta_{A,B}\equiv\zeta_B-\zeta_A$ and the overall factors of the coupling constants and the boundary value of the fields have been stripped off. These coefficients solve the e.o.m \eqref{Phi_eom} the kinetic denominators arise by rewriting the identity \eqref{Xzetaidentity},
\begin{eqn}
\left(\sum_{\ell\in Q}k_\ell\right)^2-|\vec k_Q|^2
=
2\sum_{\substack{\ell<m\\ \ell,m\in Q}}
X_{\ell m}(\zeta_m-\zeta_\ell).
\end{eqn}
To obtain the wavefunction coefficient we evaluate \eqref{eq:app_bulk_coeff} at $t=0$ and substitute
the resulting boundary expansion into
\eqref{eq:wavefunction_expansion}. Matching the coefficient
of the four ordered free waves $1234$ with the response
$i\partial_t\Phi^{\rm bulk}_{1234}|_{t=0}$ we get
\begin{eqn}\label{eq:app_psi5_recursion}
\psi_5(1,2,3,4,5)
={}&
\frac{2i\sqrt{2}(k_5-k_1-k_2-k_3-k_4)}
{\Delta_{1,2}\Delta_{2,3}\Delta_{3,4}}
+i\sqrt{2}\Bigg[
\frac{\psi_4(1,2,34,5)}{\Delta_{3,4}}
+\frac{\psi_4(1,23,4,5)}{\Delta_{2,3}}
+\frac{\psi_4(12,3,4,5)}{\Delta_{1,2}}
\Bigg]
\\
&\quad+2\Bigg[
\frac{\psi_3(1,234,5)}
{\Delta_{2,3}\Delta_{3,4}}
+\frac{\psi_3(12,34,5)}
{\Delta_{1,2}\Delta_{3,4}}
+\frac{\psi_3(123,4,5)}
{\Delta_{1,2}\Delta_{2,3}}
\Bigg].
\end{eqn}
The three $\psi_4$ terms correspond to the partitions $\tau_4,\tau_5,\tau_6$ where one boundary field is replaced by its two-wave bulk coefficient $-i\sqrt{2}/\Delta_{a,a+1}$. The three $\psi_3$ terms correspond to $\tau_1,\tau_2,\tau_3$ and contain either one three-wave coefficient or a product of two two-wave coefficients. The terms in the first line come from two contributions. The term proportional to $k_5$ arises from differentiating the subtracted homogeneous mode $e^{ik_5t}$. The remaining terms arise from applying $i\partial_t$ to the product of four free plane waves, $e^{i (k_1 + k_2 + k_3 + k_4) t}$, which produces the factor $-(k_1+k_2+k_3+k_4)$.

We now recover this expression from $\det A_5$. At five points the matrix can be constructed in the same manner as the four point case \ref{ssec:detstructure}. We remind the reader that  $\tau_j=(Q_1|\cdots|Q_s)$ is the partition labeling column $j$, and $s$ is the number of blocks in that partition. Thus both $s$ and the blocks $Q_a$ depend on $j$. For this example, the explicit  ordered partitions $\tau$'s are given as $\tau_1 = (1|234), \tau_2 = (12|34), \tau_3 = (123|4), \tau_4 = (1|2|34), \tau_5 = (1|23|4), \tau_6 = (12|3|4), \tau_7 = (1|2|3|4)$. From this the matrix can be constructed in an analogous manner resulting in a $7\times7$ matrix. To avoid a cluttering of notations we define
\begin{eqn}
P=(1234), \qquad r_j\equiv
\frac{\zeta_{Q_s}-\zeta_P}
{\prod_{a=1}^{s-1}\Delta_{Q_a,Q_{a+1}}}
\end{eqn}
where $r_j$ denotes the last row entries and are explicitly given as 
\begin{eqn}\hspace{-0.5cm}
r_1 &= \frac{\zeta_{234}-\zeta_{1234}}
{\Delta_{1,234}},
\
r_2 = \frac{\zeta_{34}-\zeta_{1234}}
{\Delta_{12,34}},
\
r_3 = \frac{\zeta_4-\zeta_{1234}}
{\Delta_{123,4}},
\
r_4 = \frac{\zeta_{34}-\zeta_{1234}}
{\Delta_{1,2}\Delta_{2,34}},
\
r_5 = \frac{\zeta_4-\zeta_{1234}}
{\Delta_{1,23}\Delta_{23,4}},
\
r_6 = \frac{\zeta_4-\zeta_{1234}}
{\Delta_{12,3}\Delta_{3,4}},
\
r_7 = \frac{\zeta_4-\zeta_{1234}}
{\Delta_{1,2}\Delta_{2,3}\Delta_{3,4}}.
\end{eqn}
Using the refinement rule in \eqref{eq:npt_refinement}, we obtain the matrix,
\begin{eqn}\label{eq:app_A5}
\setlength{\arraycolsep}{7pt}
A_5=
\begin{pmatrix}
1&0&0&
\frac{1}{\Delta_{2,34}}&
\frac{1}{\Delta_{23,4}}&
0&
\frac{1}{\Delta_{2,3}\Delta_{3,4}}
\\[7pt]
0&1&0&
\frac{1}{\Delta_{1,2}}&
0&
\frac{1}{\Delta_{3,4}}&
\frac{1}{\Delta_{1,2}\Delta_{3,4}}
\\[7pt]
0&0&1&
0&
\frac{1}{\Delta_{1,23}}&
\frac{1}{\Delta_{12,3}}&
\frac{1}{\Delta_{1,2}\Delta_{2,3}}
\\[7pt]
0&0&0&1&0&0&
\frac{1}{\Delta_{3,4}}
\\[7pt]
0&0&0&0&1&0&
\frac{1}{\Delta_{2,3}}
\\[7pt]
0&0&0&0&0&1&
\frac{1}{\Delta_{1,2}}
\\[7pt]
r_1&r_2&r_3&r_4&r_5&r_6&r_7
\end{pmatrix}.
\end{eqn}
We can now see how the lower-point terms appearing in
the perturbiner solution \eqref{eq:app_psi5_recursion}
are encoded in $A_5$. Expanding along the last column,
\begin{eqn}
\det A_5=\sum_{j=1}^{7}(A_5)_{j7}\mathcal C_{j7},
\end{eqn}
where $\mathcal C_{j7}$ denotes the cofactor. The minors simplify by expansion along rows containing
a single unit entry. For example,
\begin{eqn}
\mathcal C_{47}
&=
-\det
\begin{pmatrix}
1 & 0 & \frac{1}{\Delta_{2,34}}\\[4pt]
0 & 1 & \frac{1}{\Delta_{1,2}}\\[4pt]
r_1 & r_2 & r_4
\end{pmatrix}
=
-\left(
r_4-\frac{r_1}{\Delta_{2,34}}
-\frac{r_2}{\Delta_{1,2}}
\right)
=\frac{\psi_4(1,2,34,5)}{2k_{5\bar w}},
\end{eqn}
where the original 6$\times$6 co-factor reduces to an effective $3\times3$ matrix due to the sparse structure. Similarly,
\begin{eqn}
\mathcal C_{57}
&=\frac{\psi_4(1,23,4,5)}{2k_{5\bar w}},
\qquad
\mathcal C_{67}
=\frac{\psi_4(12,3,4,5)}{2k_{5\bar w}}.
\end{eqn}
The remaining cofactors are
$\mathcal C_{17}=-r_1$,
$\mathcal C_{27}=-r_2$,
$\mathcal C_{37}=-r_3$ and
$\mathcal C_{77}=1$.
Substituting these into the last column expansion gives
\begin{eqn}\label{eq:app_A5_expansion}
\det A_5
={}&r_7
+\frac{1}{2k_{5\bar w}}\Bigg[
\frac{\psi_4(1,2,34,5)}{\Delta_{3,4}}
+\frac{\psi_4(1,23,4,5)}{\Delta_{2,3}}
+\frac{\psi_4(12,3,4,5)}{\Delta_{1,2}}
\Bigg]-\frac{r_1}{\Delta_{2,3}\Delta_{3,4}}
-\frac{r_2}{\Delta_{1,2}\Delta_{3,4}}
-\frac{r_3}{\Delta_{1,2}\Delta_{2,3}}.
\end{eqn}
Similarly, the three point contributions arise from the first three cofactors in the last column expansion. 
Finally, using the explicit expression for $r_7$,
\begin{eqn}\label{eq:app_A5_energy}
k_{5\bar w}r_7
+\frac{k_{1\bar w}}{\Delta_{2,3}\Delta_{3,4}}
+\frac{k_{12\bar w}}{\Delta_{1,2}\Delta_{3,4}}
+\frac{k_{123\bar w}}{\Delta_{1,2}\Delta_{2,3}}
=
\frac{k_5-k_1-k_2-k_3-k_4}
{\Delta_{1,2}\Delta_{2,3}\Delta_{3,4}}.
\end{eqn}
This identity follows from
$|\vec k_Q|=\zeta_Q k_{Q\bar w}-k_{Qz}$
and spatial momentum conservation. Multiplying by $2i\sqrt{2}$ reproduces the first
line of \eqref{eq:app_psi5_recursion}. Together with the three four-point contributions, this accounts for every term in the perturbiner
solution and establishes
\begin{eqn}
\psi_5(1,2,3,4,5)
=
2i\sqrt{2}\,k_{5\bar w}\det A_5.
\end{eqn}
This demonstrates how the determinant structure maps to the perturbiner for higher point functions.

\subsection{Family by Family Ward Identities: Derivation}\label{app:WIfamily}
We give more details on the the implications of the ward identities family by family. 
We demonstrate the consequences for the Ward identities for the zeroth and the first family and discuss the role of higher families towards the end. 

\subsubsection{Zeroth Family}
 The symmetries for this are given as
\begin{eqn}
\delta_\Lambda \Phi = \tilde\Lambda(v, w), \qquad \delta_\Lambda \bar\Phi = 0
\end{eqn}
where $\square \Lambda = 0$. 
The Ward identity for this follows from \eqref{Sdvar},
\begin{eqn}\label{zerofamWI}
\Tr \int d^3 x \left[ - i \tilde \Lambda \frac{\delta \mathcal W}{\delta \varphi} - \bar\varphi \dot{\tilde\Lambda}  \right] = 0
\end{eqn}
The choice of $\tilde\Lambda$ which ensures the correct fall off at $t \to -\infty$ is given as 
\begin{eqn}
\tilde \Lambda = e^{i \sqrt{2} \a v+i\beta w} f^aT^a \equiv  e^{i \sqrt{2} \a v+i\beta w} \bm{T}
\end{eqn} 
for $\a > 0$ and arbitrary $\beta$ and $f^a$. $\tilde\L$ and its derivative at $t = 0$ are
\begin{eqn}
\Bigl. \tilde \Lambda \Bigr|_{t = 0} = e^{i \a z +i\beta w}\bm{T}, \qquad 
\Bigl.\dot{\widetilde\Lambda}\Bigr|_{t=0}
=i\alpha\left.\widetilde\Lambda\right|_{t=0} 
\end{eqn}
Plugging this into \eqref{zerofamWI}, for the first term we get
\begin{eqn}
\Tr \int d^3 x\left(  - i \tilde\Lambda \frac{\delta \mathcal W}{\delta \varphi(\vec x)} \right) &=-i \Tr\, \bm{T} \int d^3 x e^{i \vec p \cdot \vec x}  \frac{\delta \mathcal W}{\delta \varphi(\vec x)} \Bigg|_{\vec p=\a\widehat z +\b\widehat w}  
= -i \bm{T}   \Tr\, \frac{\delta \mathcal W}{\delta \varphi(\vec p)}\Bigg|_{\vec p=\a\widehat z +\b\widehat w}
\end{eqn}
The second term can also be written in a similar form 
\begin{eqn}\label{w2termincharge}
\Tr \int d^3 x \bar\varphi \dot{\tilde \Lambda} &=i    \Tr\, \bm{T} \a  \bar\varphi(-\vec p)\Bigg|_{\vec p=\a\widehat z +\b\widehat w} = i \Tr\, \bm{T}    p_z  \bar\varphi(-\vec p)\Bigg|_{\vec p=\a\widehat z +\b\widehat w}
\end{eqn}
 This term combines with the $\mathcal W_2$ term in \eqref{zerofamWI}. Hence the full contribution can be written in a suggestive form
\begin{eqn}\label{firstorderWI1}
\Tr \int d^3 x \Bigg[ - i \tilde \Lambda \frac{\delta \mathcal W_2}{\delta \varphi(\vec x)} - \bar\varphi \dot{\tilde\Lambda} \Bigg]_{\vec p = \a \widehat z+ \b \widehat w} -i \Tr\, \bm{T}\left.
\sum_{N=3}^{\infty}g^{N-2}
\frac{\delta \mathcal W_N}{\delta\varphi(\vec p)}
\right|_{\vec p=\alpha\widehat z+ \b \widehat w} = 0 
\end{eqn}
The first term above vanishes using $\mathcal W_2 \sim \Tr\int d^3 p p \varphi(\vec p) \bar \varphi(-\vec p)$ and \eqref{w2termincharge} as shown below,
\begin{eqn}
\begin{aligned}
&\Tr\int d^3x\left[
-i\widetilde\Lambda\frac{\delta \mathcal W_2}{\delta\varphi(\vec x)}
-\bar\varphi\,\dot{\widetilde\Lambda}
\right] \Bigg|_{\vec p=\alpha\widehat z +\b \widehat w}
=-i\Tr\,\bm{T}\left[
(p_z-p)\bar\varphi(-\vec p)
\right]\Big|_{\vec p=\alpha\widehat z +\b \widehat w}
=0.
\end{aligned}
\end{eqn}
where we have used that $p=\sqrt{2 p_w p_{\wb} + p_z^2}$. Thus the constraint \eqref{firstorderWI1} becomes
\begin{eqn}
\Tr\, \bm{T}\left.
\sum_{N=3}^{\infty}g^{N-2}
\frac{\delta \mathcal W_N}{\delta\varphi(\vec p)}
\right|_{\vec p=\alpha\widehat z+ \b \widehat w} = 0 
\end{eqn}
Since this is true for arbitrary $f^a$ we have the following constraint 
\begin{eqn}\label{WNconstraint}
\left.
\sum_{N=3}^{\infty}g^{N-2}
\frac{\delta \mathcal W_N}{\delta\varphi(\vec p)}
\right|_{\vec p=\alpha\widehat z+ \b \widehat w} = 0 
\end{eqn}

Hence this implies that for every wavefunction coefficient we have the constraint
\begin{eqn}\label{zerofamconst}
 \Psi_{N}(\vec p, \vec k_1, \cdots, \vec k_{N-1})\Big|_{\vec p = \a \widehat z+\b \widehat w} = 
 \Psi_{N}(\vec p, \vec k_1, \cdots, \vec k_{N-1})\Big|_{p_{\wb}=0}
 =0 , \qquad \forall\  N \geq 3.
\end{eqn}

We also emphasize that the constraint above is valid for $\a > 0$. This already shows that even the zeroth family places constraints on the wavefunction coefficients. 

\subsubsection{First and Higher Families}\label{app:first_higher_families}
\label{app:first_higher_families}
We now show how the higher families give further constraints (see \eqref{S1soft}) on the colour ordered wavefunction coefficients \eqref{eq:wavefunction_expansion}. Consider the wavefunction coefficient with the color ordering $(\vec p,1,\ldots,n)$, where $\vec p$ labels a $\varphi$ leg and $n$ labels the $\bar\varphi$ leg. The corresponding colour trace is
$\Tr(\bm T T_1\cdots T_n)$. 

The first-family Ward identity is given as,
\begin{equation}\label{firstWI}
\Tr\int d^3x\left(
-i\widetilde\Lambda\frac{\delta\mathcal W}{\delta\varphi}
-\bar\varphi\dot{\widetilde\Lambda}
+g[\Lambda,\varphi]\frac{\delta\mathcal W}{\delta\varphi}
+g[\Lambda,\bar\varphi]\frac{\delta\mathcal W}{\delta\bar\varphi}
-ig\bar\varphi[\dot\Lambda,\varphi]
\right)=0.
\end{equation}
We choose
\begin{equation}
\Lambda_1=e^{i\sqrt{2}\alpha v+i\beta w}\bm T,
\quad
\widetilde\Lambda_1
=i(\beta u+\sqrt{2}\alpha\bar w)\Lambda_1,
\qquad \alpha>0.
\end{equation}
At $t=0$, these parameters obey
\begin{eqn}
\Lambda_1|_{t=0}
&=e^{i\alpha z+i\beta w}\bm T,
\quad
\dot\Lambda_1|_{t=0}
=i\alpha e^{i\alpha z+i\beta w}\bm T,
\quad
\widetilde\Lambda_1|_{t=0}
=i\left(-\frac{\beta z}{\sqrt{2}}
+\sqrt{2}\alpha\bar w\right)
e^{i\alpha z+i\beta w}\bm T. 
\end{eqn}
Thus $p_z=\alpha$, $p_w=\beta$ and $p_{\bar w}=0$ and $\a >0$ from the Bunch-Davies condition. First, notice that the contribution of $\mathcal W_2$ in the inhomogeneous term cancels the term linear in $\bar\varphi$ in \eqref{firstWI}. Indeed, $\psi_2=-k$ and
$\dot{\widetilde\Lambda}(0,\vec k)
=ik\widetilde\Lambda(0,\vec k)$ from \eqref{gauge_falloff} give
\begin{equation}
\begin{aligned}
&\Tr\int d^3x\left(
-i\widetilde\Lambda\frac{\delta\mathcal W_2}{\delta\varphi}
-\bar\varphi\dot{\widetilde\Lambda}\right)=
i\Tr\int d^3k\,k\left[
\widetilde\Lambda(0,\vec k)\bar\varphi(-\vec k)
-\bar\varphi(-\vec k)\widetilde\Lambda(0,\vec k)
\right]=0.
\end{aligned}
\end{equation}
This cancellation removes the field-linear part of the Ward identity. Note that it does not remove the $\mathcal W_2$ terms generated by the commutators, which contribute to the $3\to2$ point relation as shown below.

Using
$z e^{i\vec p\cdot\vec x}=-i\partial_{p_z}e^{i\vec p\cdot\vec x}$
and
$\bar w e^{i\vec p\cdot\vec x}
=-i\partial_{p_{\bar w}}e^{i\vec p\cdot\vec x}$,
the remaining inhomogeneous contribution becomes
\begin{equation}\label{Lamdwdphi}
\begin{aligned}
&-i\Tr\int d^3x\,
\widetilde\Lambda_1\frac{\delta\mathcal W_{n+1}}{\delta\varphi(\vec x)} =
i\Tr\left[
\bm T\left(\frac{\beta}{\sqrt{2}}\partial_{p_z}
-\sqrt{2}\alpha\partial_{p_{\bar w}}\right)
\frac{\delta\mathcal W_{n+1}}{\delta\varphi(\vec p)}
\right]_{\substack{
p_z=\alpha,\\
p_w=\beta,\\
p_{\bar w}=0}}
=
-i\sqrt{2}\,p_z\Tr\left[
\bm T\partial_{p_{\bar w}}
\frac{\delta\mathcal W_{n+1}}{\delta\varphi(\vec p)}
\right]_{p_{\bar w}=0},
\quad n\geq2.
\end{aligned}
\end{equation}
The $p_z$ derivative vanishes due to the zeroth-family constraint holds for every $n+1\geq3$. To extract the momentum space coefficient without ambiguity, we use the momentum-conserving delta function to set
\begin{equation}
\vec k_n=-\vec p-\sum_{r=1}^{n-1}\vec k_r
\end{equation}
before taking the derivative. A similar choice is made for deriving the diagrammatic soft limits \cite{Chowdhury:2024dcy}. The $\p_{\bar w}$ derivative can then act either on the wavefunction coefficient or on the boundary source carrying $\vec k_n$. The latter contribution is proportional to $\psi_{n+1}|_{p_{\bar w}=0}=0$, and therefore vanishes due to \eqref{zerofamconst}. 
At order $g^{n-1}$, the
coefficient of the ordered trace $\Tr(\bm T T_1\cdots T_n)$ from
\eqref{Lamdwdphi} is consequently
\begin{equation}
-i\sqrt{2}\,p_z
\left.\partial_{p_{\bar w}}
\psi_{n+1}(\vec p,\vec k_1,\ldots,\vec k_n)
\right|_{p_{\bar w}=0}.
\end{equation}

We next evaluate the two commutator terms in \eqref{firstWI}. Multiplication by $\Lambda_1|_{t=0}=e^{i\vec p\cdot\vec x}\bm T$ replaces the momentum of the transformed hard leg by $\vec k_r+\vec p$. For the fixed hard cyclic ordering $(1,\ldots,n)$, the term acting on $\varphi$ therefore gives
\begin{equation}
\sum_{r=1}^{n-1}
\psi_n(\vec k_1,\ldots,\vec k_r+\vec p,\ldots,\vec k_n)
\Tr\!\left(T_1\cdots[\bm T,T_r]\cdots T_n\right).
\end{equation}
We have omitted the common factors $g^{n-1}\delta^3(\vec p+\sum_{r=1}^n\vec k_r)$. Expanding each commutator gives
\begin{equation}
\begin{aligned}
\Tr\!\left(T_1\cdots[\bm T,T_r]\cdots T_n\right)
={}&\Tr(T_1\cdots T_{r-1}\bm T T_r\cdots T_n)-\Tr(T_1\cdots T_r\bm T T_{r+1}\cdots T_n).
\end{aligned}
\end{equation}
We now extract the coefficient of \(\operatorname{Tr}(TT_1\cdots T_n)\). In this cyclic ordering, \(T\) lies between \(T_n\) and \(T_1\). The first trace in (C20) therefore contributes only for \(r=1\), where \(T\) is inserted immediately before \(T_1\). The second trace would contribute only for \(r=n\), where \(T\) is inserted immediately after \(T_n\), but this value is excluded from the sum in (C19). Thus the contribution to the selected trace from the \(\varphi\) transformations is $+\psi_n(\vec k_1+\vec p,\vec k_2,\ldots,\vec k_n).$

The $\bar\varphi$ transformation gives
\begin{equation}
\psi_n(\vec k_1,\ldots,\vec k_{n-1},\vec k_n+\vec p)
\Tr\!\left(T_1\cdots T_{n-1}[\bm T,T_n]\right).
\end{equation}
Its trace is
\begin{equation}
\begin{aligned}
\Tr\!\left(T_1\cdots T_{n-1}[\bm T,T_n]\right)
={}&\Tr(T_1\cdots T_{n-1}\bm T T_n)
-\Tr(T_1\cdots T_n\bm T)
=\Tr(\bm T T_n T_1\cdots T_{n-1})
-\Tr(\bm T T_1\cdots T_n).
\end{aligned}
\end{equation}
It therefore contributes as $-\psi_n(\vec k_1,\ldots,\vec k_{n-1},\vec k_n+\vec p)$  to the selected trace. Terms with other hard cyclic orderings cannot
contribute to this same trace. 

For example, at four points, the three commutator contributions can be displayed in full. For the ordering $(1,2,3)$, with leg $3$ being $\bar\varphi$, they are
\begin{equation}
\begin{aligned}
&\psi_3(\vec k_1+\vec p,\vec k_2,\vec k_3)
\left[\Tr(\bm T T_1T_2T_3)-\Tr(\bm T T_2T_3T_1)\right]
+\psi_3(\vec k_1,\vec k_2+\vec p,\vec k_3)
\left[\Tr(\bm T T_2T_3T_1)-\Tr(\bm T T_3T_1T_2)\right]\\
&\quad+\psi_3(\vec k_1,\vec k_2,\vec k_3+\vec p)
\left[\Tr(\bm T T_3T_1T_2)-\Tr(\bm T T_1T_2T_3)\right],
\end{aligned}
\end{equation}
and the coefficient of $\Tr(\bm T T_1T_2T_3)$ is precisely
\begin{equation}
\psi_3(\vec k_1+\vec p,\vec k_2,\vec k_3)
-\psi_3(\vec k_1,\vec k_2,\vec k_3+\vec p).
\end{equation}

Finally, using $\dot\Lambda_1=ip_z\Lambda_1$, the last term of
\eqref{firstWI} is
\begin{equation}
\begin{aligned}
-ig\Tr\int d^3x\,\bar\varphi[\dot\Lambda_1,\varphi]
&=gp_z\int d^3x\,e^{i\vec p\cdot\vec x}
\Tr\!\left(\bar\varphi[\bm T,\varphi]\right)
=gp_z\int d^3x\,e^{i\vec p\cdot\vec x}
\left[\Tr(\bm T\varphi\bar\varphi)
-\Tr(\bm T\bar\varphi\varphi)\right].
\end{aligned}
\end{equation}
There are only two fields in this expression and contributes
$+p_z$ to the coefficient of $\Tr(\bm T T_1T_2)$ for $n=2$, and
vanishes at every higher multiplicity.

Combining all terms, the ordered Ward identity is
\begin{equation}\label{eq:first_family_ordered_WI}
\begin{aligned}
i\sqrt{2}\,p_z
\left.\partial_{p_{\bar w}}
\psi_{n+1}(\vec p,\vec k_1,\ldots,\vec k_n)
\right|_{p_{\bar w}=0} = 
\Big[
\psi_n(\vec k_1+\vec p,\vec k_2,\ldots,\vec k_n)
-\psi_n(\vec k_1,\ldots,\vec k_{n-1},\vec k_n+\vec p)
\Big]_{p_{\bar w}=0}
+p_z\delta_{n,2}.
\end{aligned}
\end{equation}
Dividing by $i\sqrt{2}p_z$ gives \eqref{S1soft}. 

\end{document}